\documentclass[aps,pra,preprint,12pt,a4paper]{revtex4-2}

\usepackage{amssymb}
\usepackage{amsmath}
\usepackage{bm}
\usepackage{natbib}
\usepackage{graphics,graphicx}
\usepackage{epsfig}
\usepackage{latexsym}
\usepackage{epic,eepic,graphicx,amssymb,amsmath,indentfirst}
\usepackage{bm}
\usepackage{graphicx}
\usepackage[utf8]{inputenc}
\usepackage[T1]{fontenc}
\usepackage{natbib}
\usepackage{color}
\usepackage[colorlinks={true}]{hyperref}
\hypersetup{citecolor={blue}, filecolor={blue}, linkcolor={blue}, urlcolor={blue}}
\newcommand{\bra}[1]{\langle #1 \vert}
\newcommand{\ket}[1]{\vert #1 \rangle}

\usepackage{amsmath,amsfonts,amssymb,latexsym,mathrsfs,amsthm,amstext, bezier,amscd,indentfirst}
\usepackage{subcaption}
\usepackage[font={normalsize,it}]{caption}

\begin{document}

\title{\bf Formation of oriented polar molecules with a single shaped pulse}

\author{Murilo D. Forlevesi}
\email{murilo.deliberali@unesp.br }
\author{Edson Denis Leonel}
\email{edson-denis.leonel@unesp.br }
\affiliation{Departamento de Física, UNESP - Universidade Estadual Paulista, Rio Claro SP, 13506-900, Brazil}
\author{Emanuel F. de Lima}
\email{eflima@ufscar.br}
\affiliation{Departamento de F\'isica, Universidade Federal de S\~ao Carlos (UFSCar)\\ S\~ao Carlos SP, 13565-905, Brazil}
\date{\today}

\begin{abstract}
We explore the possibility of forming a oriented polar molecule directly from a pair of colliding atoms. The process comprises the photoassociation and vibrational stabilization along with the molecular orientation. These processes are driven by a single time-dependent, linearly polarized control field and proceeds entirely within the electronic ground state, leveraging the presence of a permanent dipole moment. The control field is found by means of an optimal quantum control algorithm with a single target observable given by the restriction of the orientation operator on a subset of bound levels. We consider a rovibrational model system for the collision of the O + H atoms and solve directly the time-dependent Schrodinger equation. We show that the optimized field is capable of enhancing the molecular orientation already induced by the photoassociation and vibrational stabilization thus yielding oriented polar molecules that can be useful for many applications.
\end{abstract}


\maketitle
\newpage


\section{Introduction}

The coherent control of quantum systems via externally applied fields is fundamental for atomic and molecular physics \cite{10.1063/1.449767,10.1063/1.450074,F29868202423}. Quantum optimal control stands as a broad framework to determine control fields aimed to reach a desired goal, usually translated as the expectation value of an observable \cite{10.1063/1.471215,Brif_2010,10.1063/1.458438,doi:https://doi.org/10.1002/9783527639700.ch5}. In fact, the use of temporally and spectrally shaped pulses to steer the dynamics has enabled the realization of a wide range of control objectives, including state-selective excitation, reaction pathway control, and quantum gate implementation \cite{Chang2015,PhysRevLett.126.163202,10.1063/1.3185565,PhysRevLett.129.050507}. Among the various control tasks, two problems of enduring importance are photoassociation of atoms into deeply bound molecular vibrational states and the orientation or alignment of polar molecules along a laboratory-fixed axis.


In photoassociation, colliding atoms are led to form a molecule by the action of an external laser field \cite{PhysRevA.63.013412,PhysRevA.107.023114,PhysRevA.96.043417,PhysRevLett.114.233003}. There are several reasons to pursue the control of photoassociation, among them, the obtainment of ultracold molecules from pre-cooled atoms, which can allow for application in quantum computation and in precise measurement of physical constants \cite{koch2012}. Moreover, photoassociation stands as fundamental step towards the effective control of chemical reactions \cite{RevModPhys.78.483}. In most of the photoassociation schemes, the first step consists of an induced transition to a bound rovibrational level of an excited electronic state by a cw-laser or a pump pulse in the visible or ultraviolet domain. This is followed by a induced or spontaneous transition to the ground electronic state, the dumping process. It is desirable for many applications that the second step leaves the molecule in low lying vibrational levels, where the molecule is stable against dissociation \cite{PhysRevA.98.053411,PhysRevA.109.013315,PhysRevA.95.013411,Lyu2019}. This vibrational cooling procedure is referred to as vibrational stabilization \cite{PhysRevA.82.043437,zt28-q4pn,deAlmeida_2019,PhysRevLett.109.183001,PhysRevA.99.053428,Reich_2013}.

A different approach to photoassociation is the use of infrared or microwave radiation to form a molecule directly in the ground state \cite{KOROLKOV199796,KOROLKOV1996604,DELIMA200648,PhysRevA.78.063417,PhysRevLett.99.073003}. This alternative route is possible for the collision of distinct atomic species due to the associated dipole moment coupling free and bound rovibrational levels of the electronic ground state. The advantage of this scheme is that it does not depend either on the structure or lifetimes of excited states. Furthermore, this process can create polar molecules, which are of special interest due to its long range dipole-dipole interaction \cite{Ulmanis2012,PhysRevLett.92.033004}. It is worth mentioning that other methods are also available to produce vibrationally excited molecules from colliding atoms, e.g., magneto-association, which must often be combined with a vibrational stabilization process \cite{PhysRevA.88.063418, PhysRevLett.124.253401,Warner_2023}.


The ability of producing spatially aligned or oriented gas-phase polar molecules can be decisive for a large range of applications in chemistry and physics, including high-harmonic generation, molecular tomography, laser pulse compression and electron diffraction \cite{RevModPhys.91.035005,GAMBHIR2025112598,RevModPhys.75.543,PhysRevLett.74.4623,D3CP05592B,ZHOU2025141880}. Molecular alignment refers to the adjustment of the molecular axis along a well-defined spatial direction, without distinguishing between opposite orientations. By contrast, molecular orientation denotes the preferential direction of the molecular dipole toward one hemisphere, thereby breaking inversion symmetry \cite{D3CP03115B,10.1063/1.4973773}. Although high degree of orientation of polar molecules can be achieved by the application of strong DC fields, its use can be detrimental in altering desired experimental conditions \cite{PhysRevA.103.L041103,PhysRevA.87.013429}. Alternatively, the molecular orientation under field-free conditions is often pursued. In this case, the orientation occurs after the interaction of a pulse, which causes the superposition of rotational states allowing for the recurrence of the orientation. Recently, it has been proposed the combination of two-color slow turn-on and rapid turn-off laser pulse to induce rotational transitions. It has been demonstrated that this method yields highly orientated molecules considering only a limit number of rotational levels \cite{D3CP03115B}. In the same context, a theoretical framework employing a pulse sequence was developed to achieve high orientation by controlling a subset of rotational states \cite{PhysRevResearch.7.L012049}.  

The relation between alignment and photoassociation has been reported early. It has been found that the pump photoassociation pulse induce a superposition of rovibrational levels in the excited electronic state, which lead to a partial field-free alignment of the excited-state molecule, affecting the dumping process \cite{10.1063/1.4929388}. It has also been shown that a chirped field can generate partially aligned molecules via two-photon photoassociation \cite{PhysRevA.96.053613}. In these cases, the molecular alignment were not specifically controlled, but rather a consequence of the photoassociation process, while the molecular orientation were not pursued.


In the present work, we tackle the problem of combining photoassociation, vibrational stabilization and orientation in a single control objective. In particular, we pose the following question: is it possible, starting from two colliding atoms, to form a oriented molecule in low vibrational levels by means of a single pulse? To this end, we define as a target observable the restriction of the orientation operator to a set of low-lying vibrational bound levels. For the numerical simulations, we consider a model system for the collision of O + H  with a permanent dipole coupling to a linearly polarized IR pulse \cite{DELIMA2011267}. The control pulse is calculated by means of the two-point boundary-value quantum control paradigm (TBQCP) accelerated monotonic scheme \cite{PhysRevE.82.026703,PhysRevA.84.031401}. 

The paper is organized as follows: Section II is reserved to the general formulation of the model. Subsection II-A presents the perturbed and unperturbed Hamiltonian, the Morse potential and the distribution of the initial condition of the system. In subsection II-B, we describe the time-dependent Schrodinger equation and the method to solve it. In II-C, we describe the optimization algorithm and the cost functional that will be maximized. In section III, we discuss the results obtained. In III-A we present the control of the rovibrational system only for molecular photoassociation. Section III-B is dedicated to the control of molecular orientation. In III-C we present the simultaneous control of two processes: photoassociation and molecular orientation. Finally, section IV is reserved to conclusions of the study.

\section{Model System}

Consider a collision of two atoms of distinct species in the electronic ground state in the presence of an electric field generated by a linearly polarized infrared laser pulse. In this situation, the magnetic quantum number $m$ is conserved and for simplicity we consider $m=0$. The laser interacts with the colliding atoms through the associated dipole moment and we can write the Hamiltonian of this system as \cite{deLima2015},

\begin{equation} \label{Eq.1}
H(r,t)=H_0(r)+H_1(r,t)
\end{equation}
with, 
\begin{equation} \label{Eq.2}
H_0(r)=\frac {- \hbar^2} {2 M_r}\frac{1}{r} \frac{\partial^2}{\partial r^2} + \frac{\hat {\textbf{L}}} {2M_r r^2} + V(r), 
\end{equation}
where $r= |\vec {r}|$ is the internuclear separation, $M_r$, $\hat{\textbf{L}}$ and $V(r)$ are, respectively, the reduced mass, the angular momentum operator and the internuclear potential energy. The interaction Hamiltonian $H_1(r,t)$ is given by,

\begin{equation} \label{Eq.3}
H_1(r,t)=-\vec{\mu}(r) \cdot \vec{u}(t)= - \mu(r)u(t)\cos(\theta),
\end{equation}
with  $\mu(r)$, $u(t)$ and $\theta$ being, respectively, the dipole momentum function, the electric field and the angle between the field polarization axis and the dipole moment. The dipole moment function has the form \cite{KOROLKOV199796} ,
\begin{equation} \label{Eq.4}
\mu(r)= qr \exp{\bigg(- \frac{r}{r_d} \bigg)}
\end{equation}
where $q$ is the effective charge and $r_d$ sets the range of the dipole interaction. The potential energy $V(r)$ is a Morse function truncated at long range L by an infinite barrier \cite{deLima_2008},
\begin{equation} \label{Eq.5}
V(r)=
\begin{cases}
V_M(r), ~~~~ & {\rm for}~~ r\leq L  \\
\infty, ~~~~ & {\rm for}~~ r>L

\end{cases}
\end{equation}

The Morse potential is given by
\begin{equation} \label{Eq.6}
V_M(r)=D_e[{\rm e}^{-2\alpha(r-r_e)}-2{\rm e}^{-\alpha(r-r_e)}],
\end{equation}
where $D_e$ is the well depth at the equilibrium position $r_e$, and $\alpha^{-1}$ is the potential range. The infinite barrier is located sufficiently far away from the potential well and from the effect of the dipole, such that it should not influence the system dynamics. Convergence of the numerical results with increasing values of $L$ is checked \textit{a posteriori}. 

We consider the basis of the eigenstates of $ H_0$ which have the form,
\begin{equation}
    \bra{r} kl \rangle=\frac{\phi_{kl}(r)}{r}Y^{0}_l(\theta),
\end{equation}
where $Y^{0}_l(\theta)$ are the spherical harmonic with the magnetic quantum number $m=0$ and rotational quantum number $l=0,1,2,\ldots$. The radial wavefunctions $\phi_{kl}(r)$ are solutions of the eigenvalue equation,

\begin{equation}
    \left[\frac{\hbar^2}{2M_r}\frac{d^2}{dr^2}+\frac{\hbar^2 l(l+1)}{2M_r r^2}+V(r)\right]\phi_{kl}(r)=E_{kl}\phi_{kl}(r),
\end{equation}
where $E_{kl}$ is the energy corresponding to the rovibrational state $\ket{kl}$. For each value of the angular momentum, the spectrum of $H_0$ is composed by a finite number of bound discrete states with radial functions $\phi_{\nu l}(r)$ and by an infinite number of unbound states with radial functions $\phi_{n l}(r)$ (the unbound states are also discrete due to the presence of the infinite barrier). We denote the number of bound levels for a given $l$ by $N_b^l$.  For practical numerical calculations, the maximum number of levels for any orbital angular momentum is truncated at some large value $N$, so that the number of unbound states is $K^l=N-N_b^l$. In order to notationally distinguish the scattering and bound states, the latter are labeled by the Greek letter $\nu=0,1,...,N_b^l-1$, whereas the unbound are denoted by roman letter $n=1,2,\ldots,K^l$. 
Also for the sake of numerical calculations, we consider that the system has a maximum value for the angular momentum, $l=0,\ldots, l_{max}$. Thus, the total number of rovibrational levels considered is $N_D=N (l_{max}+1)$.

In this study, we consider $l=0$ collisions, described by an initial Gaussian wavepacket written as,
\begin{equation} \label{eq:initial_wp}
\psi(r,0) = \frac{\xi(r)}{r}Y_{0}^0(\theta)
\end{equation}
where $ \xi(r)$ is the Gaussian Wavepacket
\begin{equation} \label{Eq.8}
\xi(r) = \bigg ( \frac{2}{\pi d^2}  \bigg)^{1/4} e^{ik_0r-(r-r_0)^2/d^2},
\end{equation}
where $r_0$, $hk_0 < 0$ and $d$ are, respectively, the center of the wavepacket, the relative momentum and the width of the wave packet. The initial center of the wavepacket $r_0$ is considered sufficiently far from the influence of the potential function and the dipole, that is, $ V_M (r_0) \approx  0$ and $r_0\gg r_d$. 

Consider the time-dependent Schrodinger equation that governs the system dynamics,

\begin{equation} \label{Eq:tdse}
i\hbar \frac{\partial}{\partial t} \psi (r,t)=[H_0(r)+H_1(r,t)]\psi(r,t).
\end{equation}
At any time, we assume the wave function can be written in terms of the truncated basis of eigenstates of $H_0$, i.e., 

\begin{equation} \label{Eq:expan_basis}
\psi(r,t)= \frac{1}{r} \bigg[\sum^{N_b^l-1}_{\nu=0} \sum^{l_{max}}_{l=0}a_{\nu l}(t) \phi_{\nu l}(r)Y_l^0(\theta) + \sum^{K^l}_{n=1} \sum^{l_{max}}_{l=0}b_{n l}(t) \phi_{n l}(r)Y_l^0(\theta) \bigg],
\end{equation}
where the expansion coefficients $a_{\nu l}(t)$ and $b_{m l}(t)$  are, respectively, associated with the bound state $\phi_{\nu l}$ and the scattering state $\phi_{ml}$ with the normalization $\sum^{N_b^l-1}_{\nu=0} \sum^{l_{max}}_{l=0}|a_{\nu l}(t)|^2 + \sum^{K^l}_{n=1} \sum^{l_{max}}_{l=0}|b_{n l}(t)|^2 =1 ~~\forall t \in [0,t_f] $.

The expansion coefficients of the initial state function $\psi(r,0)$ given by Equation \eqref{eq:initial_wp} are determined from,

\begin{equation} \label{Eq.11}
a_{\nu 0}(0)= \int^L_0 \phi_{\nu l}(r)\xi(r)dr, ~~~~ \nu=0,1,...,N_b^l-1
\end{equation}
and
\begin{equation} \label{Eq.12}
b_{n 0}(0)= \int^L_0 \phi_{n l}(r)\xi(r)dr,  ~~~~ n=1,...,K^l.
\end{equation}
We chose the initial condition such that all $a_{\nu 0}$ are essentially zero (to machine precision), representing an completely unbound initial state.

Substituting the expansion \eqref{Eq:expan_basis} in the Schrodinger equation \eqref{Eq:tdse}, we can write the equation for the coefficients in matrix form,
\begin{equation} \label{Eq.13}
i \hbar \frac{d}{dt} \Psi (t) = [\boldsymbol{H_0} - \boldsymbol{\mu} u(t) ] \Psi (t),
\end{equation}
where the time-dependent state $\Psi(t)$ is a  column  vector such that,

\begin{equation}
    (\Psi(t))_{i}=\begin{cases}
         a_{\nu l}(t), & \text{for  } i=lN+1+\nu, \\
       b_{nl}(t), &\text{for  } i=lN+N_b^l+n, 
    \end{cases}
\end{equation}
where $i=1,\ldots N(l_{max}+1)$. $\boldsymbol{H_0}$ is a diagonal matrix containing the rovibrational energies in the corresponding order,
\begin{equation}
    ( \boldsymbol{H_0})_{ii}=\begin{cases}
         E_{\nu l}, & \text{for  }  i=lN+1+\nu, \\
      E_{nl}, &\text{for  } i=lN+N_b^l+n, 
    \end{cases}
\end{equation}
and the symmetric matrix of the permanent dipole moment $\boldsymbol{\mu}$ is given by,
\begin{equation} \label{Eq.14}
(\boldsymbol{\mu})_{ij}=\begin{cases}
    \bra{\phi_{\nu l}}\mu\ket{\phi_{\nu' l'}}\alpha_{ll'},  \text{ for  } i=lN+1+\nu, \text{and } j=l'N+1+\nu' \\
    \bra{\phi_{\nu l}}\mu\ket{\phi_{n l'}}\alpha_{ll'},  \text{ for  } i=lN+1+\nu, \text{and } j=l'N+N_b^{l'}+n,\\
    \bra{\phi_{n l}}\mu\ket{\phi_{n' l'}}\alpha_{ll'},  \text{ for  } i=lN+N_b^{l}+n, \text{and } j=l'N+N_b^{l'}+n',
\end{cases}
\end{equation} 
with $(\boldsymbol{\mu})_{ij}=(\boldsymbol{\mu})_{ji}$, $\alpha_{ll'}=\left[p_{l}\delta_{ll'+1}+p_{l+1}\delta_{ll'-1}\right]$ and \cite{10.1063/1.466081} 
\begin{eqnarray}
p_{l}=\bra{Y_{l+1}^0}\cos(\theta)\ket{Y_{l}^0}=\frac{l+1}{\sqrt{(2l+1)(2l+3)}}.
\end{eqnarray}

To solve Equation \eqref{Eq.13}, the corresponding propagator $U(t,0)$ can be written in terms of the sequence of propagators $U(t+\Delta t,t)$ with respect to a small time step $\Delta t$, thus, we can transform the propagator expression into a second-order split-operator \cite{DEVRIES199195}, 

\begin{equation} \label{Eq.15}
U(t+\Delta t,t) \approx e^{-i\boldsymbol{H_0} \Delta t/2 \hbar} e^{i\mu u(t) \Delta t/\hbar} e^{-i \boldsymbol{H_0}\Delta t/2 \hbar} 
\end{equation} 
this operator leads to the following state evolution $| \Psi (t+\Delta t) \rangle = U(t+\Delta t,t) | \Psi (t) \rangle$. To evaluate the exponential in Equation \eqref{Eq.15}, we diagonalize the dipole matrix $\boldsymbol{\mu}$, since $\boldsymbol{H_0}$ is already diagonal.

\section{The Optimal Control Problem}

We set the performance measurements as functionals $J\{u(t)\}$ of the control field $u(t)$ which should be maximized. In the case we are interested only in photoassociation with a given angular momentum $l$, we can consider $J\{u(t)\}$ as the expectation value of the projection operator in a set $S$ of bound vibrational levels of the molecule \cite{PhysRevA.78.063417},
\begin{equation} \label{Eq.16}
J\{u(t)\}=\bra{\psi(t_f)}P_B^l\ket{\psi(t_f)},  
\end{equation}
with $P_B^l= \sum^{}_{\nu\in S}\ket{\nu l} \bra{\nu l}$.

On the other hand, to maximize only molecular orientation, we should consider the expectation value of the operator $\cos(\theta)$ \cite{10.1063/1.4973773},
\begin{equation}
    J\{u(t)\}=\bra{\psi(t_f)}\hat{A}\ket{\psi(t_f)}, 
\end{equation}
$\hat{A}=\cos(\theta)$.

In the present work, the ultimate goal is to maximize both photoassociation and molecular orientation. To this end, we can consider $J\{u(t)\}$ as the expectation value of the operator defined by the restriction of the operator $\hat{A}$ on a set of bound levels \cite{CohenTannoudji1991},

\begin{equation} \label{Eq.17}
\hat{B}=P_B \hat{A} P_B,
\end{equation}

Thus, the objective functional that we maximize through electric field optimization is described by,

\begin{equation} \label{Eq.18}
J\{u(t)\}=  \langle \psi(t_f) |  \hat{B}| \psi(t_f) \rangle,
\end{equation}
with $P_B=\sum_lP_B^l$.


The control field $u(t)$ can be efficiently designed based on a family of monotonically convergent iterative methods referred to as the two-point boundary-value quantum control paradigm (TBQCP) \cite{PhysRevE.82.026703}.
Consider a desired target operator $\boldsymbol{\hat{O}}$ with eigenstates $\ket{k}$. We start by considering an initial guess field $u^{(0)}(t)$. In the first iteration, the control field is adapted according to

\begin{equation} \label{Eq.19}
u^{(1)}(t)=u^{(0)}(t)+\eta s(t) f^{(1)}_{\mu}(t)
\end{equation}
and for the next iterations,
\begin{equation} \label{Eq.20}
u^{(j+1)}(t)=u^{(j)}(t)+\eta s(t) [2f^{(j+1)}_{\mu}(t)-f^{(j)}_{\mu}(t)], ~~~~j=1,2,3,...
\end{equation}
where $s(t)\geq 0$ is an envelope function for the control pulse (intended to enforce smooth turn-on and turn-off) and $\eta \geq 0$ is a properly chosen constant, which sets the pace of the iterations. $f_{\mu}(t)$ can be interpreted as the gradient of the functional with respect to the field and is given by,
\begin{equation} \label{Eq.21}
f^{(j+1)}_\mu (t) = - \frac {2}{\hbar} {\rm Im} \bigg \lbrace \sum_k \sigma_k \bra{ \psi^{(j+1)}(t)} \chi^{(j)}_k (t) \rangle \times \bra{\chi^{(j)}_k(t) } \mu \ket{\psi^{(j+1)} (t)} \bigg \rbrace 
\end{equation}
where the wavefunction $\ket{\psi(t)}$ is propagated forward in time with the evolution operator associated with Eq. \eqref{Eq:tdse}. The time interval $[0:t_f]$ is divided into $M$ grid points $t_0=0,t_1,t_2,...,t_{M-1},t_M=t_f$ evenly spaced at $\Delta t$ such that

\begin{equation} \label{Eq.22}
\vert \psi^{(j+1)}(t_i) \rangle=U_{j+1}(t_i,t_i-\Delta t) \vert \psi^{(j+1)}(t_i-\Delta t) \rangle, ~~~~\vert \psi^{(j+1)}(0) \rangle \equiv  \vert \psi(0) \rangle ~~~~i=1,2,...,M
\end{equation}
and the states $\ket{\chi_k^{(j)}}$ are propagated backwards from the eigenstates of the observable, 
\begin{equation} \label{Eq.23}
\vert \chi^{(j)}_k(t_i) \rangle=U_{j}(t_i,t_i+\Delta t) \vert \chi^{(j)}_k(t_i+\Delta t) \rangle, ~~~~\vert \chi_k (t_f) \rangle \equiv  \vert k \rangle ~~~~i=0,1,...,M-1
\end{equation}

The propagators $U_{j+1}(t_i,t_i - \Delta t) $ and $U_{j}(t_i,t_i + \Delta t) $  are, respectively, associated with the control field $u^{(j+1)}(t)$ (being updated on the fly at the current $(j+1)$th iteration) and the control field $u^{(j)}(t)$ (at the previous jth iteration). At each new iteration, the Eq. \eqref{Eq.23} is integrated backward numerically and the resultant $\vert \chi^{(j)}_k(t_i) \rangle, ~~ i=M-1,M-2,..,2,1,0$, are stored for computing the function $f^{(j+1)}_ {\mu}(t_i)$ in Eq. \eqref{Eq.21}, while propagating the state $\vert \Psi (t_i) \rangle $ forward in time.

\section{Results and Discussion}

The Morse potential parameters for the OH molecule in Eq.\eqref{Eq.6} are $D_e=5.42$eV ($\approx 43763 {\rm cm}^{-1}$) and $\alpha^{-1}=0.445$\AA, while the reduced mass is $m_r=0.94$amu. For these parameters, the Morse oscillator has 22 bound states for $l=0$, that is, $N_b^{0}=22$. we use the parameters $q=1.634{\rm e}$ and $r_d=0.6$\AA for the dipole function in Eq. \eqref{Eq.4} \cite{KOROLKOV199796}. For the numerical calculations, we consider that the number of unbound level is truncated in $N=172$ and $L=48$\AA. In all control problems, the trial field is described by
\begin{equation} \label{Eq.24}
u^{(0)}(t)=s(t)V\sin (\omega_c t )
\end{equation}
where the envelope function is set to 
\begin{equation} \label{Eq.25}
s(t)=  \begin{cases}
\sin^2(\pi t /t_f), ~~~~ & \text{for}~~ 0\leq t \leq t_f  \\
0, ~~~~ & ~~ \text{otherwise,}

\end{cases}
\end{equation}
with $V= 1063 {\rm MV/cm}^{-1}$ and $\omega_c=360 {\rm cm}^{-1}(\approx 10.8ps^{-1})$.

\begin{figure}[h]
 \centering
     \includegraphics[width=13cm]{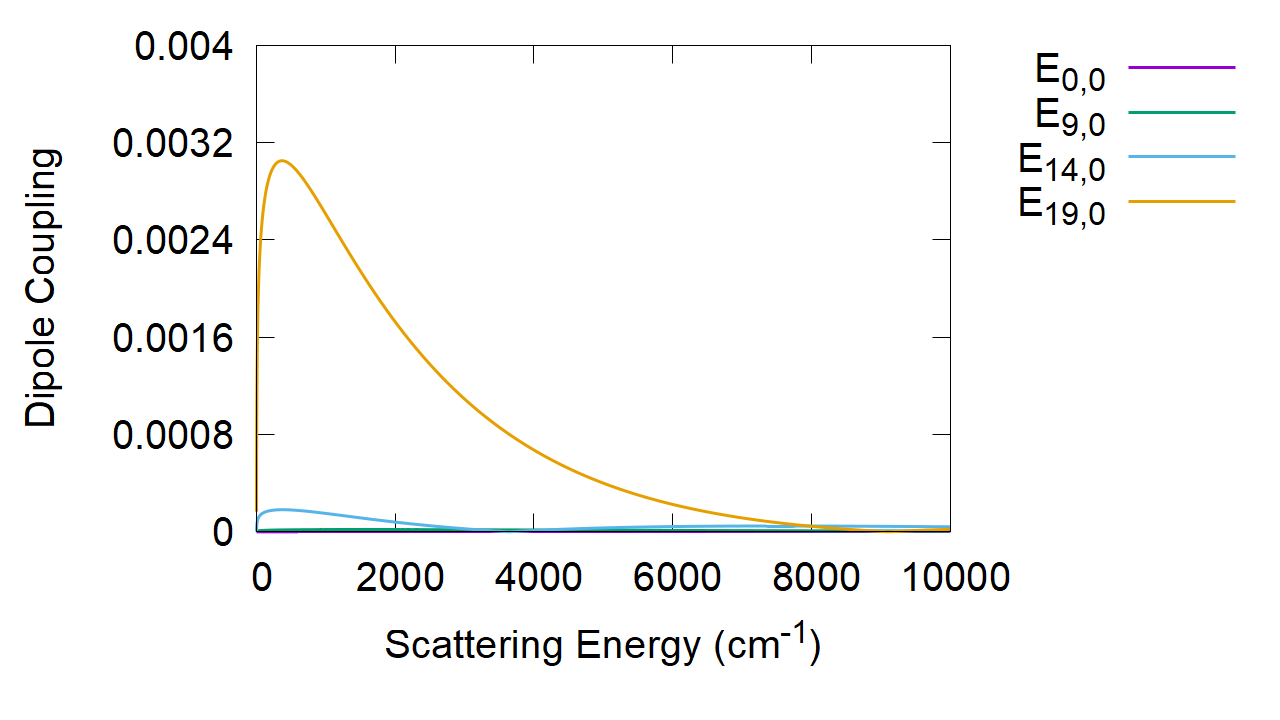}
     \caption{Absolute value of the dipole Coupling between scattering energies of $l=1$ and some bound energies levels of $l=0$, $|\bra{\phi_{\nu 0}}\mu\ket{\phi_{n1}}|$ for $\nu=0$, $9 $, $14$ and $19$.}\label{Dipolo}
\end{figure}    

Figure \ref{Dipolo} shows the dipole couplings of the scattering states of $l=1$ to some bound levels of $l=0$, $\bra{\phi_{\nu 0}}\mu\ket{\phi_{n1}}$. The coupling from the scattering states to the bound state $\nu=19$ is larger for the scattering energies in the interval $[0:7000]$, but it is considerably smaller for the vibrational state $\nu=14$ with a weak coupling occurring around the scattering energies $[0:2000]$. For $\nu=9,0$ and  $\nu=0$, the coupling is negligible. Thus, in order to bring the system to the ground vibrational level, it is first necessary to populate the higher vibrational levels and then to proceed with the vibrational stabilization. Further, due to the rotational selection rules $l\rightarrow l\pm1$, this stabilization process leads naturally to a population of several rotational states.

\subsection{Molecular Orientation}

We start considering the control of molecular orientation only, such that the initial population is concentrated in the ground vibrational and rotational state $\ket{00}$. The final time of evolution is fixed in $t_f=1.7$ps. Since the objective is to maximize the molecular orientation of the system, our target operator is written as $\boldsymbol {\hat{O}}=\hat{A}= \cos(\theta)$ and the cost functional is,

\begin{equation} \label{Eq.35}
J=  \langle \Psi(t_f) |\boldsymbol{\hat{O}}| \Psi(t_f) \rangle =\langle \Psi(t_f)  | \cos (\theta) | \Psi(t_f) \rangle . 
\end{equation}.

\begin{figure} [hb!]
 \centering
     \includegraphics[width=13cm]{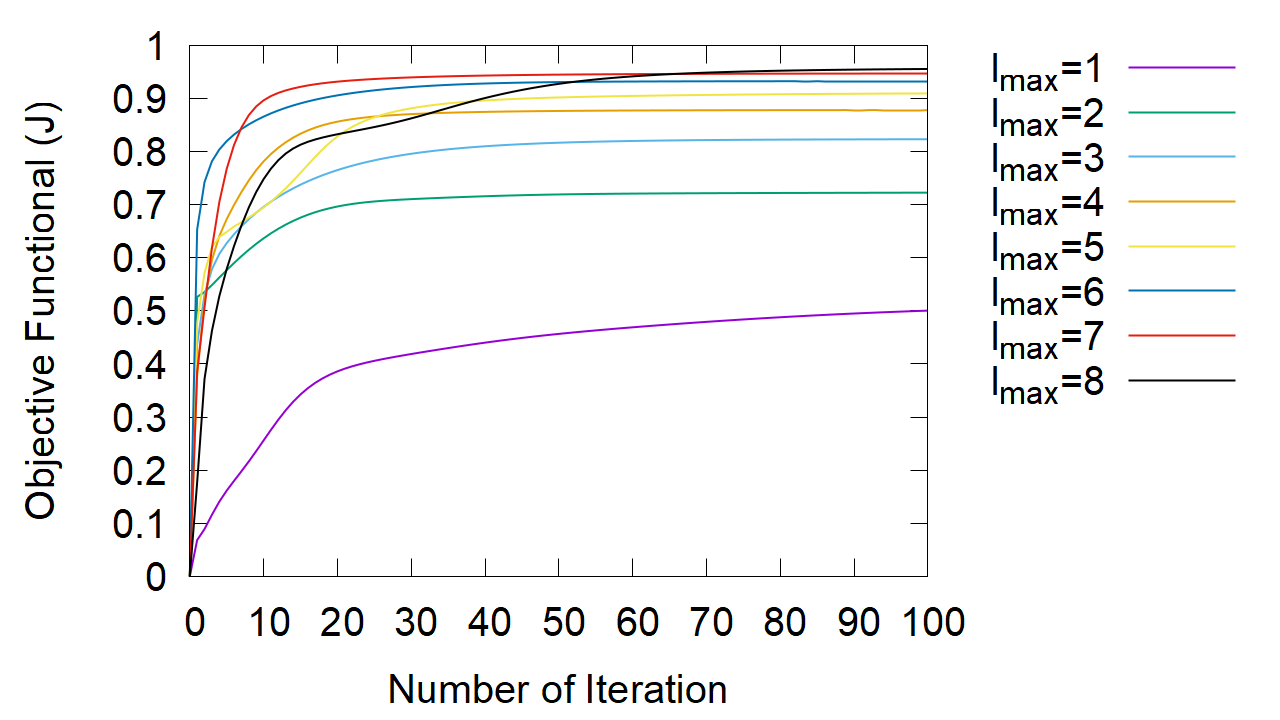}
     \caption{Molecular orientation, $\bra{\psi(t_f)} \cos (\theta) \ket{\psi(t_f)}$  at each iteration of the TBQCP algorithm for different values of the maximum angular momentum $l_{max}=1,...,8$.}\label{Fig2}
\end{figure}

Figure \ref{Fig2} shows the iterations of the TBQCP scheme, where the values of the cost functional are calculated considering the system with different maximum rotation levels $l_{max}$. For $l_{max}=1$, the cost functional reaches $J\sim 0.50$ in the 100th iteration, for $l_{max}=2$ the maximum value is $J\sim 0.72$ in the 80th iteration, for $l_{max}=3$ the functional reaches $J\sim 0.82$ in the 60th iteration, for $l_{max}=4$ it reaches $J\sim 0.88$ around the 40th iteration. Using $l_{max}=5$ the TBQCP method yield $J\sim 0.91$, for $l_{max}=6$, $J\sim 0.92$, for $l_{max}=7$, $J\sim 0.94$ and for $l_{max}=8$, $J\sim 0.96$. The variation of the value of $J$ in the end of the iterations for $l_{max}\in(1,4)$ is $\sim0.38$ and for $l_{max}=[4:8]$ is $\sim0.08$. These results show that as more rotational levels are included, more closely we can approach to perfect orientation, in agreement with \cite{D3CP03115B}. However, above $l_{max}=4$, the maximum possible value of orientation increases very slowly with $l_{max}$. Henceforth, we consider $l_{max}=4$ to balance the orientation yield with the computational cost. 

\begin{figure} [h]
 \centering
 \begin{subfigure}[h]{10cm}
     \centering
     \includegraphics[width=\textwidth]{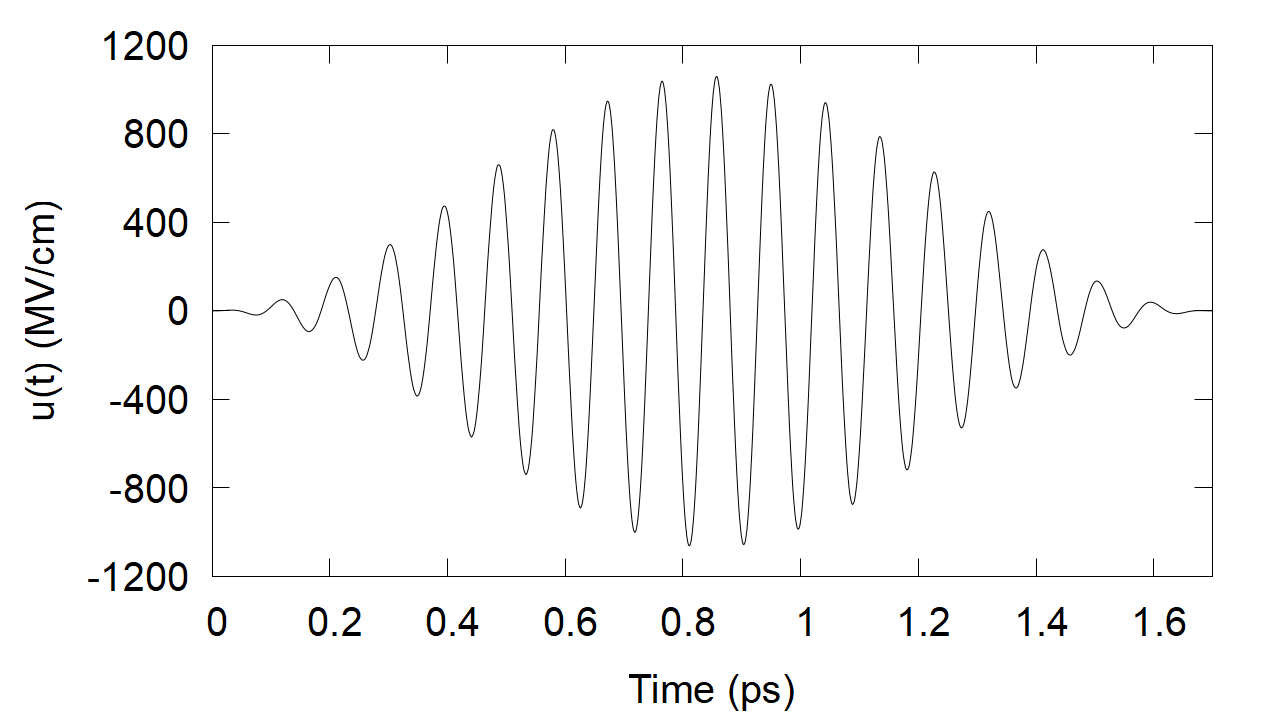}
     \caption{}
     \label{Fig3a}
 \end{subfigure}
 \hfill
 \begin{subfigure}[h]{10cm}
     \centering
     \includegraphics[width=\textwidth]{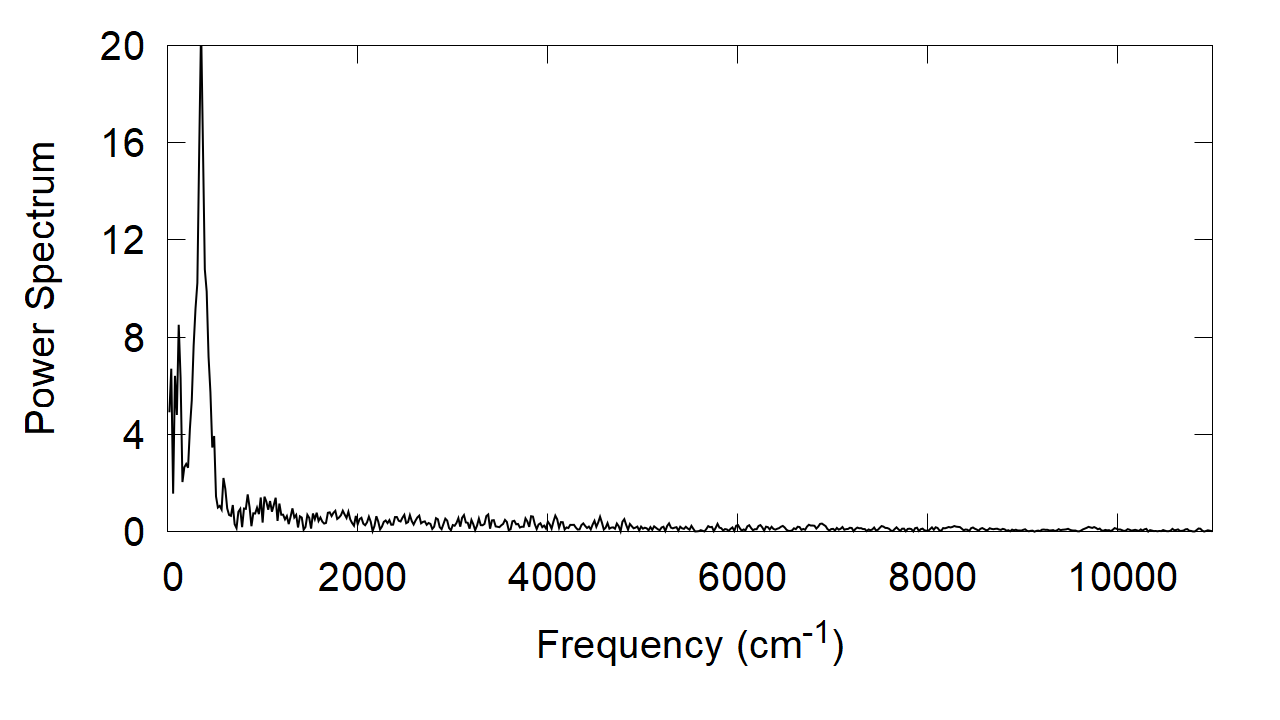}
     \caption{}
     \label{Fig3b}
 \end{subfigure}
 \hfill
 \begin{subfigure}[h]{10cm}
     \centering
     \includegraphics[width=\textwidth]{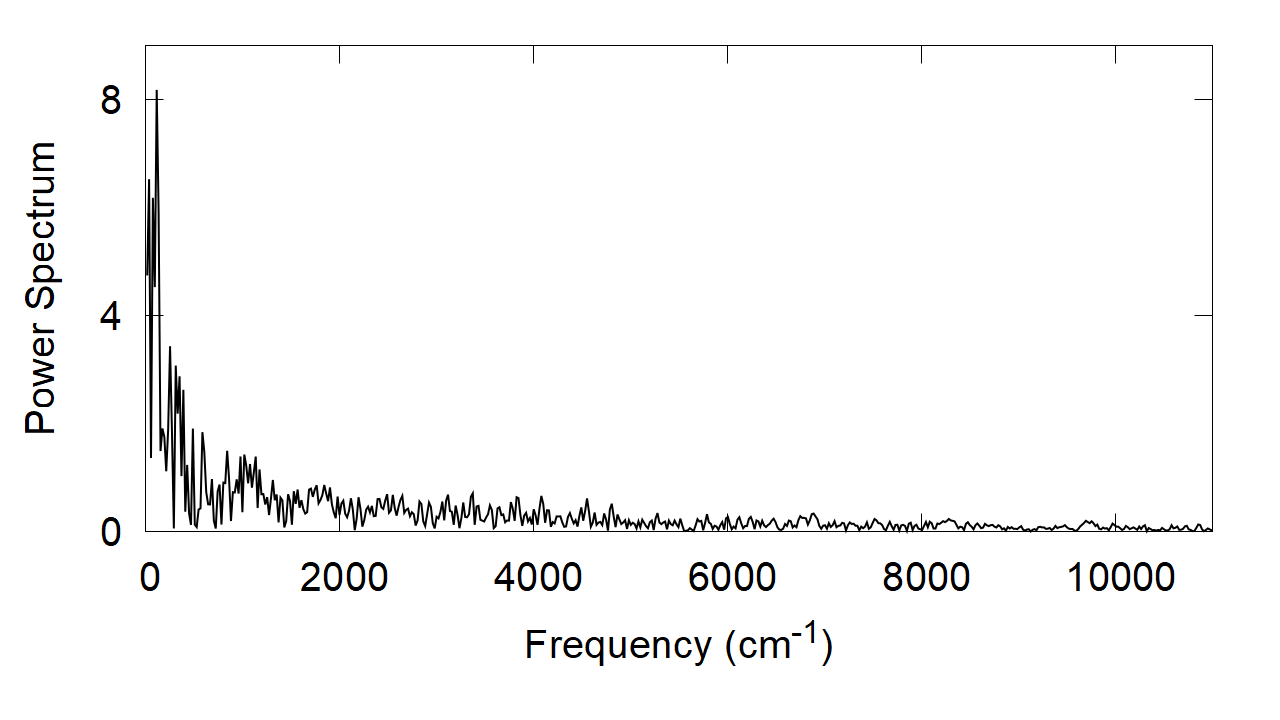}
     \caption{}
     \label{Fig3c}
 \end{subfigure}
 \hfill
    \caption{(a) Electric field resulting in the 100th iteration of the optimization algorithm. (b) The Power spectrum of the optimal field. (c) Difference in Power spectrum between trial and optimized field }\label{Fig3}
\end{figure}

Panel (a) of Figure \ref{Fig3} shows the optimized electric field at the 100th iteration and panel (b) show its power spectrum for $l_{max}=4$. The optimized field is composed by the frequency of the initial trial field $\omega_c=360$, which is the most predominant one, but there is also a region of lower frequencies with values in the interval $\omega\in(0,360)$. Frequencies higher than $\omega_c$, $\omega\in (361,11000)$, also appears in the spectrum but with considerable smaller amplitude. To visualize the frequencies that are most important for maximizing the molecular orientation, Fig.~\ref{Fig3c} shows the amplitude difference, in module, between the power spectrum of the trial and optimized field. The relevant change between transforms is concentrated in the frequency range of $\omega\in(0,11000)$, specially in to a set of low frequencies in the interval $\omega\in(0,1163)$. This result shows that lower frequencies are the most altered during the optimization of the electric field, that is, they are the most relevant, with $\omega=197.5$ having the greatest influence.

\begin{figure} [h]
 \centering
 \begin{subfigure}[h]{11cm}
     \centering
     \includegraphics[width=\textwidth]{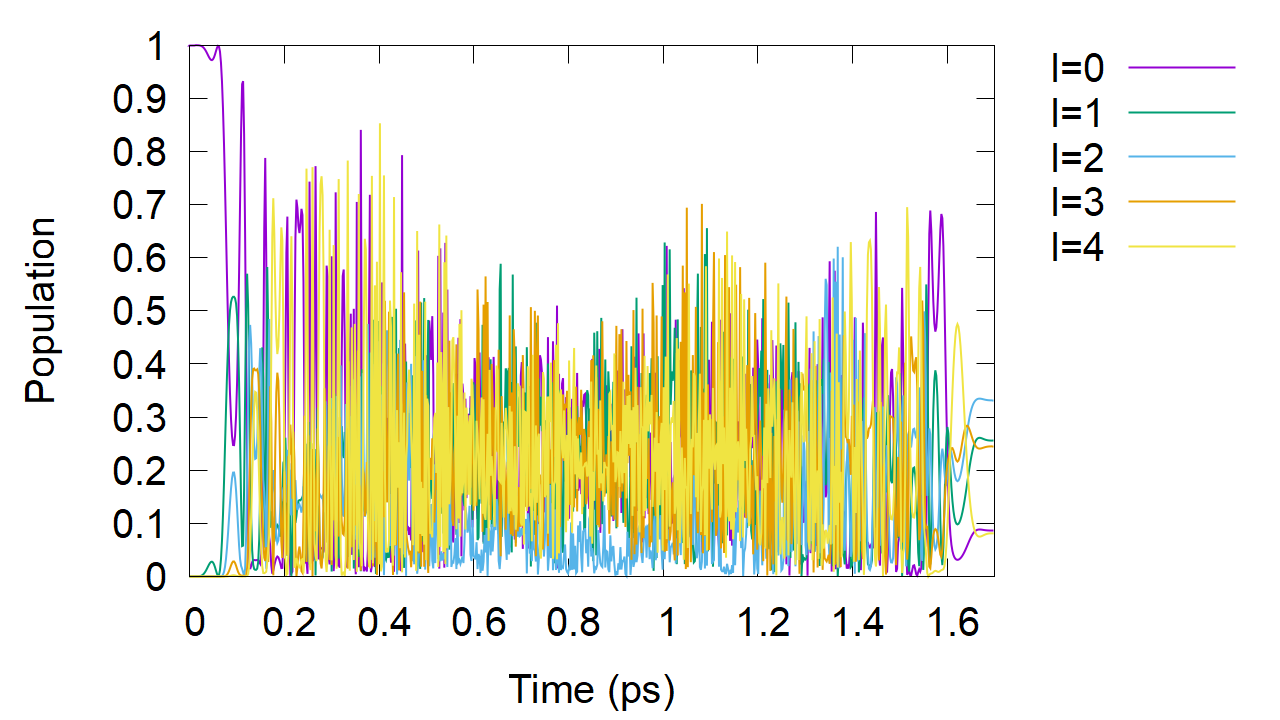}
     \caption{}
     \label{Fig5a}
 \end{subfigure}
 \hfill
 \begin{subfigure}[h]{11cm}
     \centering
     \includegraphics[width=\textwidth]{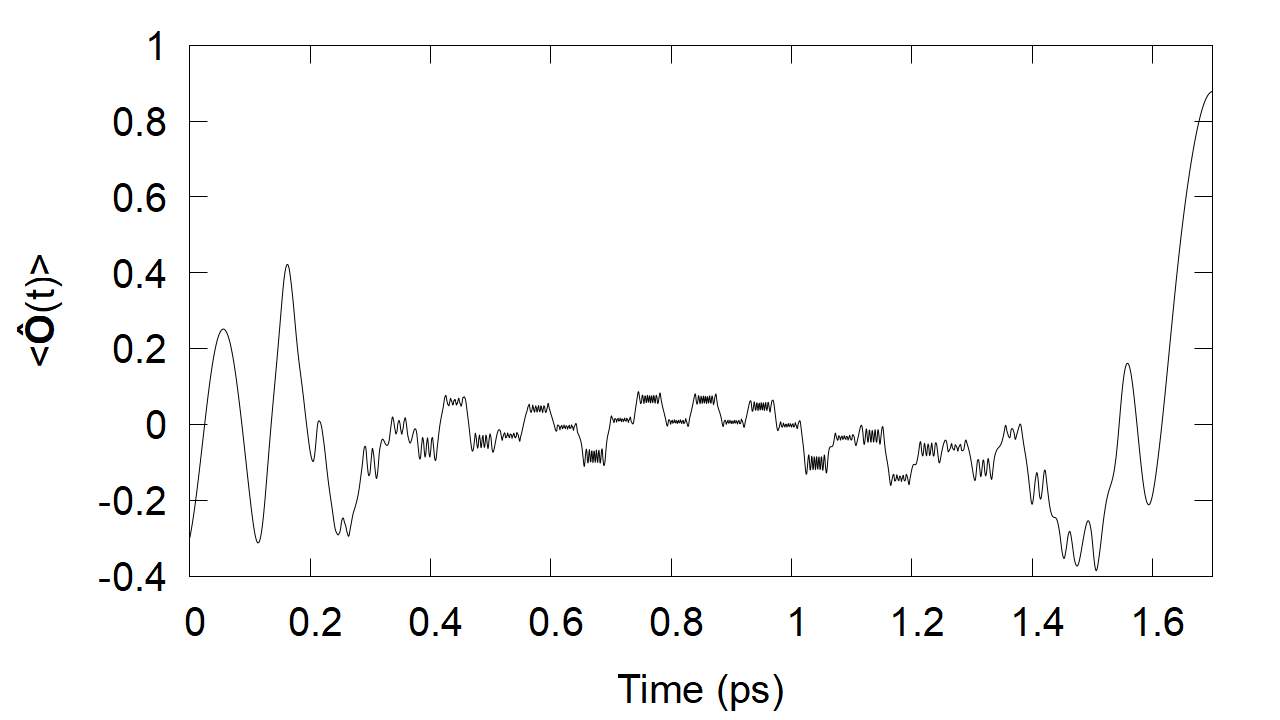}
     \caption{}
     \label{Fig5b}
 \end{subfigure}
 \hfill
    \caption[width=0.5\textwidth]{(a) Population dynamics of the in rotational levels from $l=0$ to $l=4$, with $\nu=0$. (b) Expectation value of operator $\boldsymbol{\hat O}= \cos(\theta)$ in time}\label{Fig5}

\end{figure}

The details of the population dynamics at each rotation level of the system are shown in panel (a) of Figure \ref{Fig5}, for $t=0$ the initial population is in $l=0$ (purple line), between $t=0.05$ and $t=1.6$ the population has a complex dynamic distributing between all rotation levels, in $t=1.7$ the population is organized in your optimal form. The optimal molecular orientation occurs when the population distribution is as follows, $\sim 0.09$ for $l=0$ (purple line) and for $l=4$ (yellow line); $\sim 0.26$ for $l=1$ (green line); $\sim 0.32$ for $l=2$ (blue line); and  $\sim 0.24$ for $l=3$ (orange line), note that the population tends to concentrate at intermediate rotation levels $(l=1,2,3)$. This study was also carried out for the molecule in excited states; the population distribution between the rotational levels is the same. For the optimized electric field, the Figure \ref{Fig5b} shows the expectation value of $\boldsymbol{\hat O}$, that is, the behavior of orientation of the system over time. In the interval $t\in(0,0.2)$, the orientation $\langle \boldsymbol{\hat O} (t) \rangle$ oscillates in the interval $(-0.3,0.4)$. For $0.2<t\leq1.4$, there are several plateaus with small oscillations with small positive and negative values of the orientation. Finally, for $t>1.4$, the mean value of the orientation oscillates a few times and reaches the optimal value of $0.87$ at $t=1.7$. This means that at the end of the application of the electric field, the population is distributed in its optimal configuration in the rotation levels of the system. 
\clearpage

\subsection{Photo-association}

Here, we consider only the control of photoassociation and the Gaussian wavepacket in Eq.\eqref{eq:initial_wp} is set as the initial state. Our first objective is to drive the population to the ground vibrational subspace $\nu=0$, without restricting the value of the angular momentum, except for its maximum value. The target observable is $\boldsymbol{\hat{O}}=\sum^{4}_{l=0}P_B^l=\sum^{4}_{l=0} \ket{0l} \bra{0l}$ and the target observable is given by, 

\begin{equation} \label{Eq.26}
J=  \langle \psi(t_f) |\boldsymbol{\hat{O}}| \psi(t_f) \rangle  =\sum^{4}_{l=0} |\bra{0l} \psi(t_f) \rangle |^2.
\end{equation}

The trial electric field is given by Eq.\eqref{Eq.24}, but several tests revealed that to control the photoassociation of rovibrational systems it is necessary to use fields with longer durations compared to the case of molecular orientation. Thus, we set $t_f=120$ps. 

\begin{figure}[h]
    \centering
    \includegraphics[width=11cm]{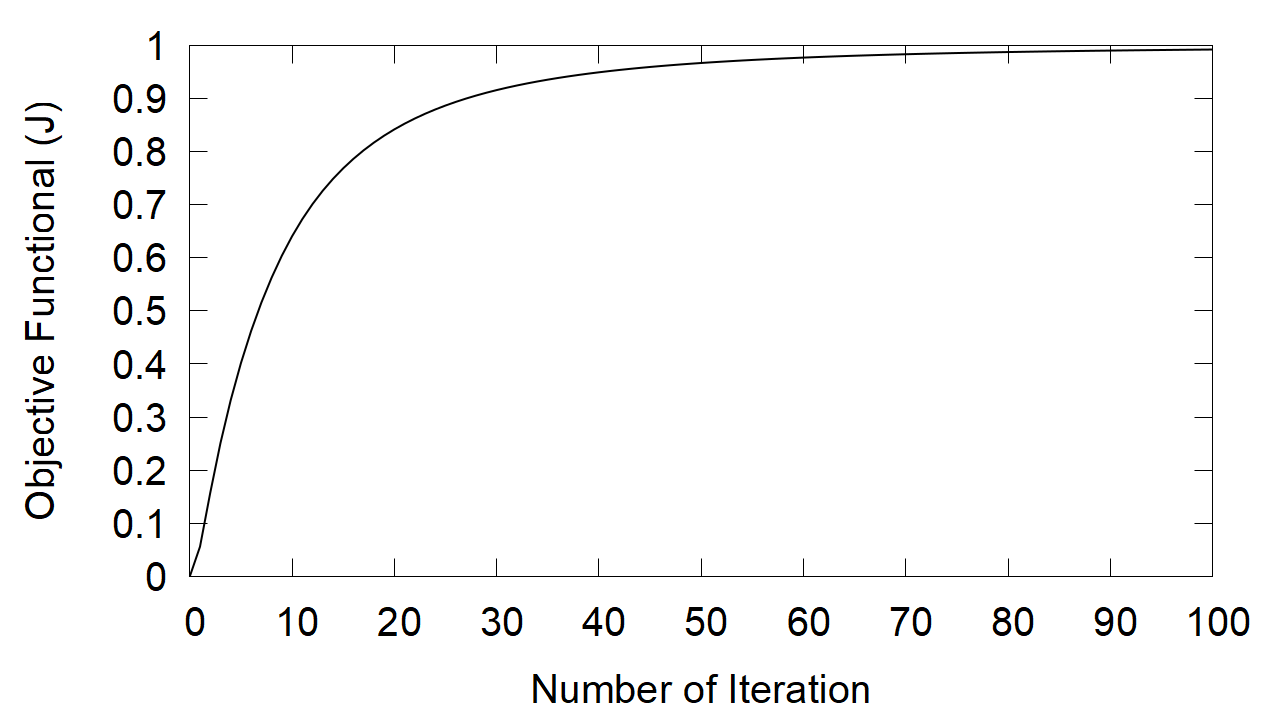}
    \caption{Expectation value of the objective functional $\boldsymbol{\hat{O}}=\sum^{4}_{l=0} \ket{0l} \bra{0l}$ for each iteration of the TBQCP algorithm for $l_{max}=4$.}\label{Fig7}
\end{figure}

\begin{figure} [h]
 \centering
 \begin{subfigure}[h]{11cm}
     \centering
     \includegraphics[width=\textwidth]{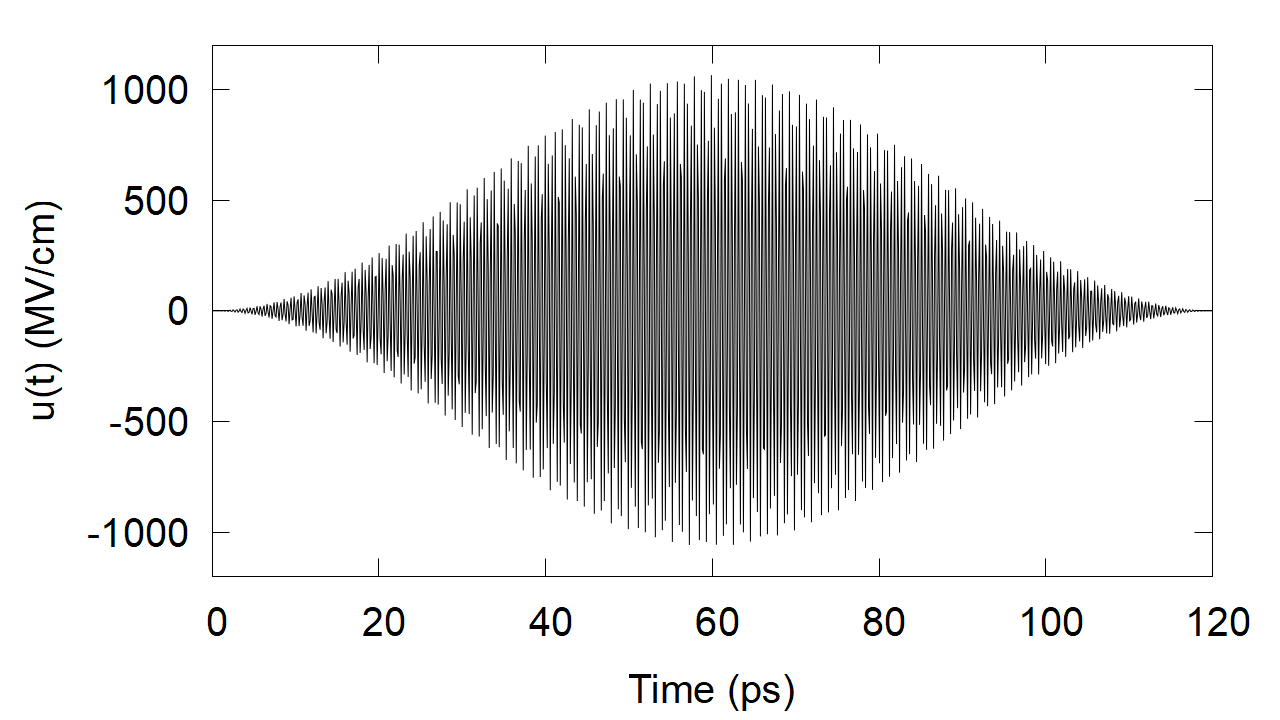}
     \caption{}
     \label{Fig;Fig1}
 \end{subfigure}
 \hfill
 \begin{subfigure}[h]{11cm}
     \centering
     \includegraphics[width=\textwidth]{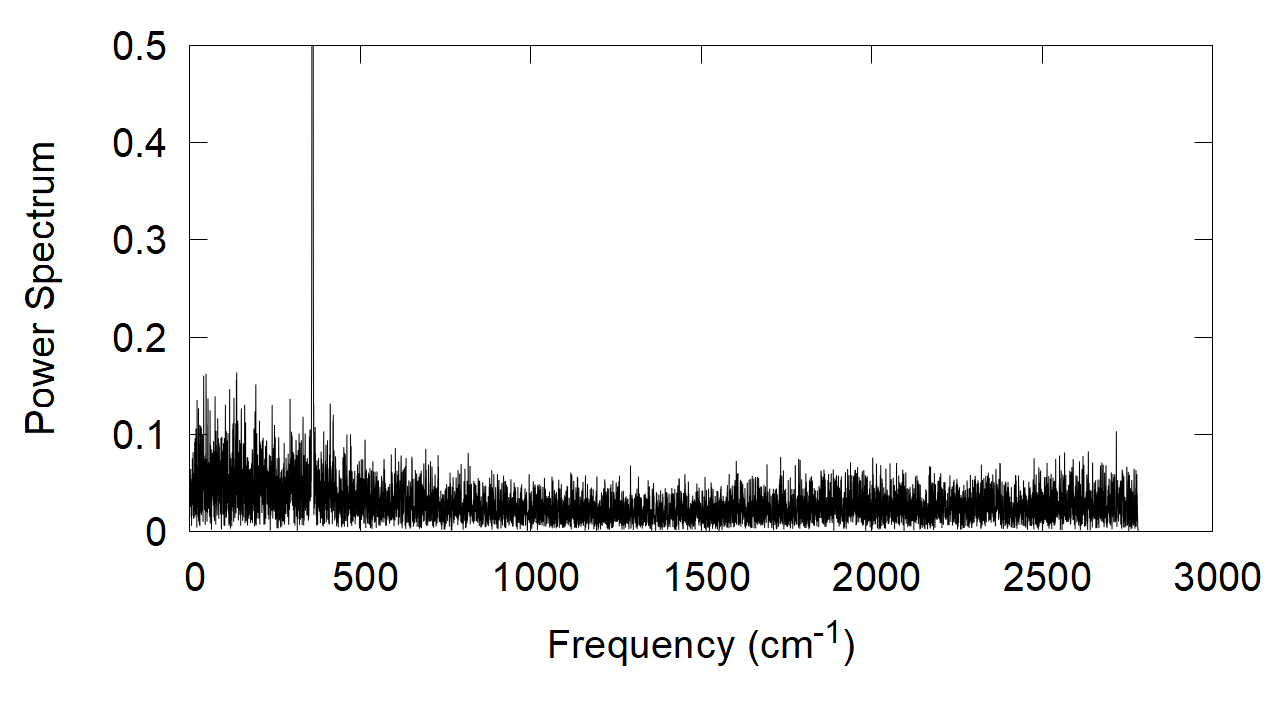}
     \caption{}
 \end{subfigure}
 \hfill
    \caption{(a) Electric field resulting in the 100th iteration of the optimization algorithm. (b) The power spectrum of the optimized field.}\label{Fig8}

\end{figure}

Figure \ref{Fig7} shows the value of the target observable at each TBQCP iteration. The functional start with $J=0$ and increases rapidly to $J=0.5$ in the 9th iteration and $J=0.8$ in the 19th iteration. From this iteration, the increasing in $J$ is slower, reaching $J\approx1$ in the 100th iteration. In comparison with the molecular orientation control, the TBQCP algorithm is able to achieve higher yields. Figure \ref{Fig8} shows the shape of the optimized electric field and its respective power spectrum. The optimized field is composed by frequencies with values that vary between $\omega=0$ and $\omega=2743$, with the frequency of the trial field, $\omega_c$, being the predominant.
\newpage

\begin{figure} [h]
 \centering
 \begin{subfigure}[h]{11cm}
     \centering
     \includegraphics[width=\textwidth]{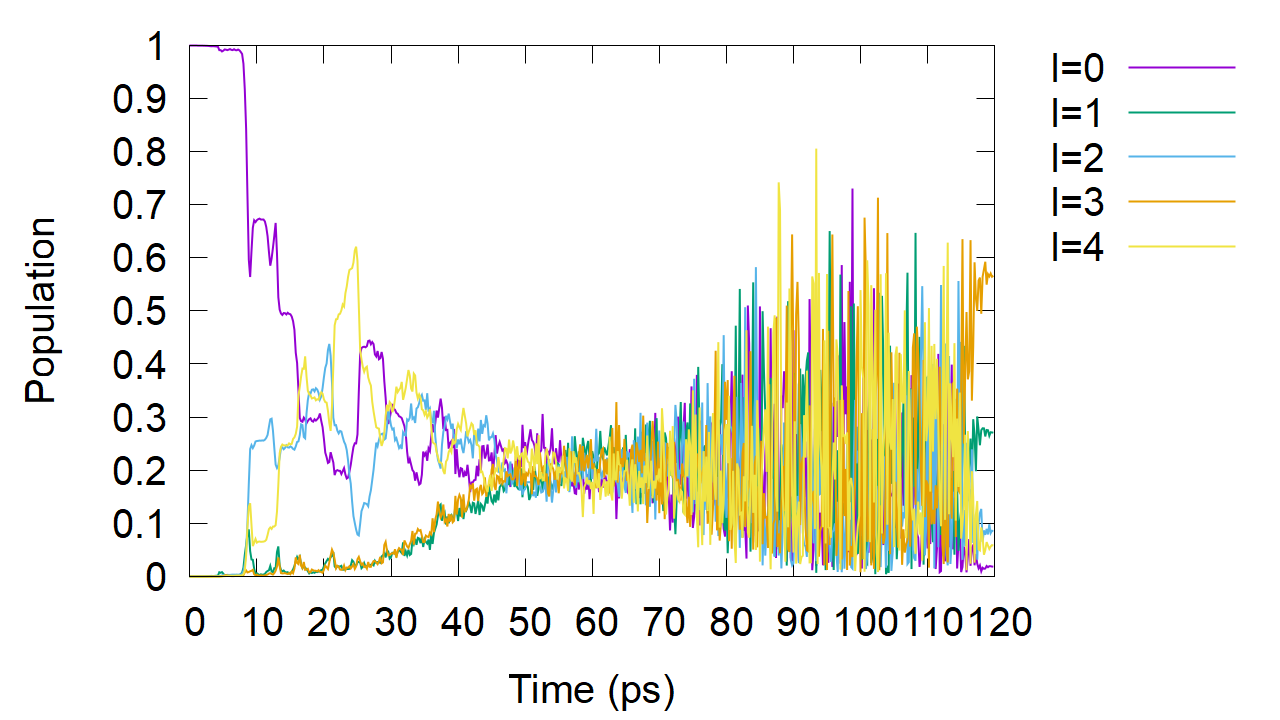}
     \caption{}
     \label{Fig9a}
 \end{subfigure}
 \hfill
 \begin{subfigure}[h]{11cm}
     \centering
     \includegraphics[width=\textwidth]{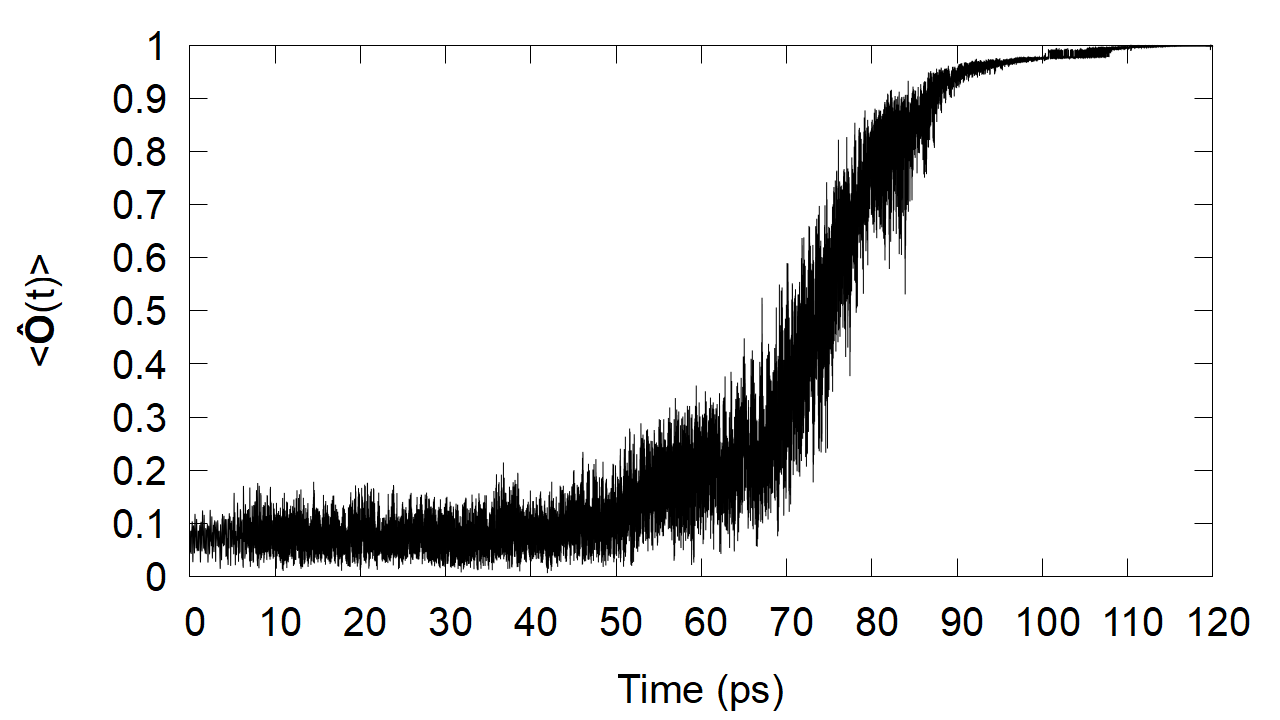}
     \caption{}
     \label{Fig9b}
 \end{subfigure}
 \hfill
    \caption{(a) Dynamic population for rotational levels $l=0,1,2,3,4$. (b) Expected value of operator $ \boldsymbol{\hat O}=\sum_{l=0}^4 {P}_B^l$  in time}\label{Fig9} 
    
\end{figure}

Figure \ref{Fig9a} shows the population dynamics of the rotational levels of the system. In this plot, the population of all vibrational levels are summed for each value of $l$. For the initial time, $t=0$, the population is all at $l=0$. In $t=5$ it begins to distribute among the other rotation degrees. In the interval $t\in(40,60)$, the population for each value of $l$ has a small variation between $0.2$ and $0.3$. For $t\in(60,115)$, this variation increases, remaining between $0$ and $0.8$. At $t=118$, the population begins to stabilize, so that at the end of the laser iteration, it is distributed as follows: $\sim 0.02$ in $l=0$; $\sim 0.27$ in $l=1$; $\sim 0.08$ in $l=2$; $\sim 0.57$ in $l=3$; and $\sim 0.06$ in $l=4$. This distribution shows that photoassociation in the  produces some degree of orientation. In fact, we obtain $\bra{\psi(t_f)}\cos(\theta)\ket{\psi(t_f)}=0.48$, which is in agreement with previous studies \cite{10.1063/1.4929388,PhysRevA.96.053613}.

Figure \ref{Fig9b} shows the expected value of the operator ${P_B}$ over time for the optimized electric field. From $t=0$ to $t=50$, the value of $P_B$ varies between 0 and 0.2, meaning that only one-fifth of the initial population occupies the $\nu=0$ subspace. In the interval $t\in(50,90)$, the mean value of the operator increases from $0.2$ to $0.9$. At the end of the action of the control field, about $0.98$ of the  population is in the $\nu=0$ rovibrational levels.

\begin{figure} [h]
 \centering
 \begin{subfigure}[h]{8cm}
     \centering
     \includegraphics[width=\textwidth]{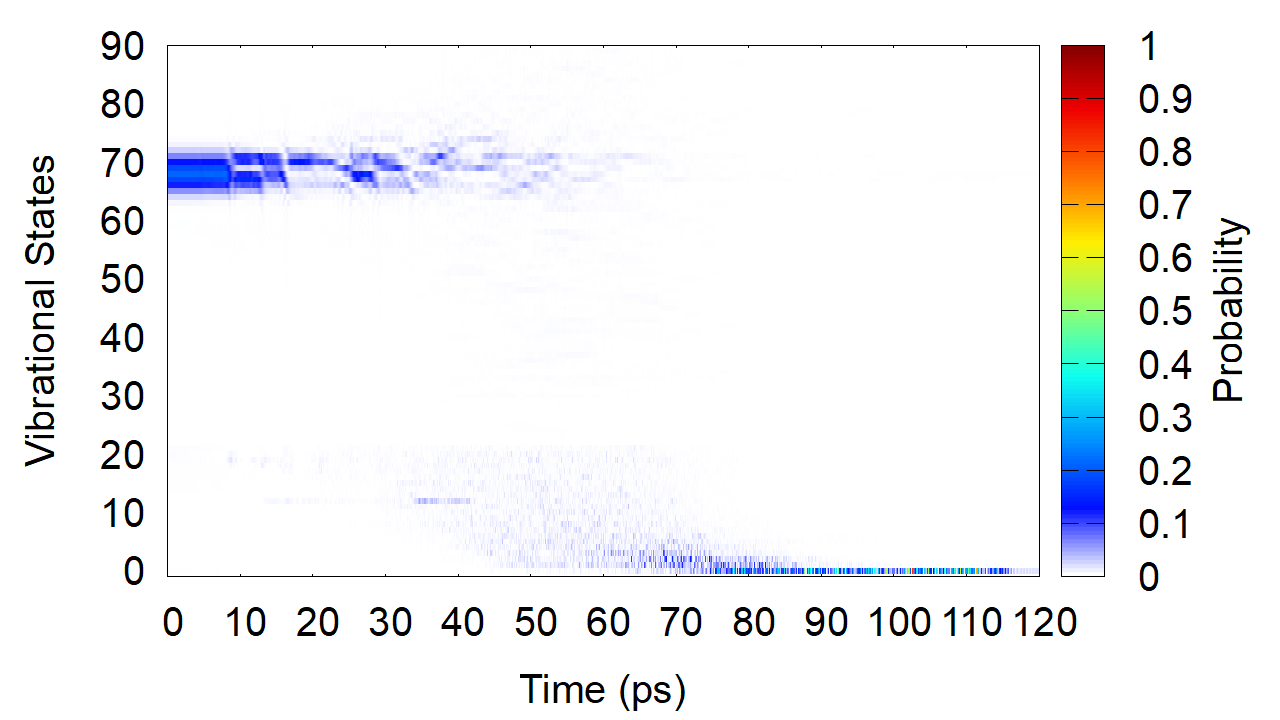}
     \caption{}
     \label{Fig10a}
 \end{subfigure}
 \hfill
 \begin{subfigure}[h]{8cm}
     \centering
     \includegraphics[width=\textwidth]{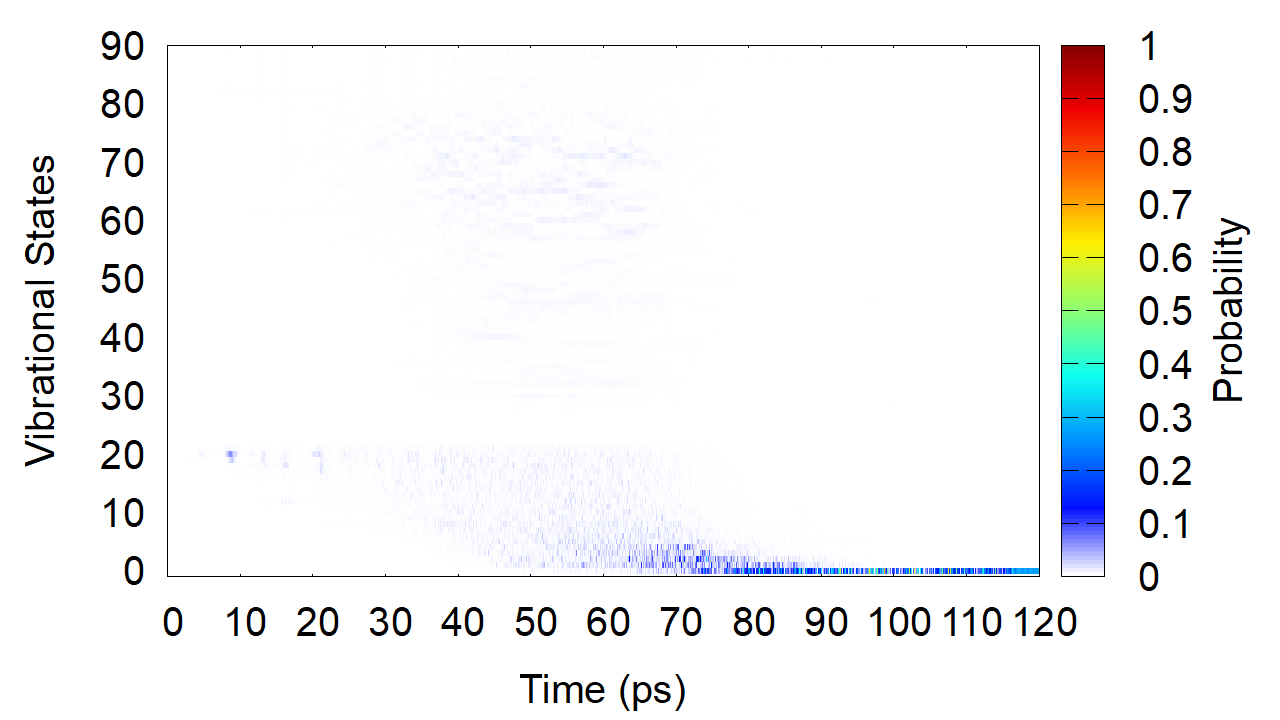}
     \caption{}
     \label{Fig10b}
 \end{subfigure}
 \hfill
 \begin{subfigure}[h]{8cm}
     \centering
     \includegraphics[width=\textwidth]{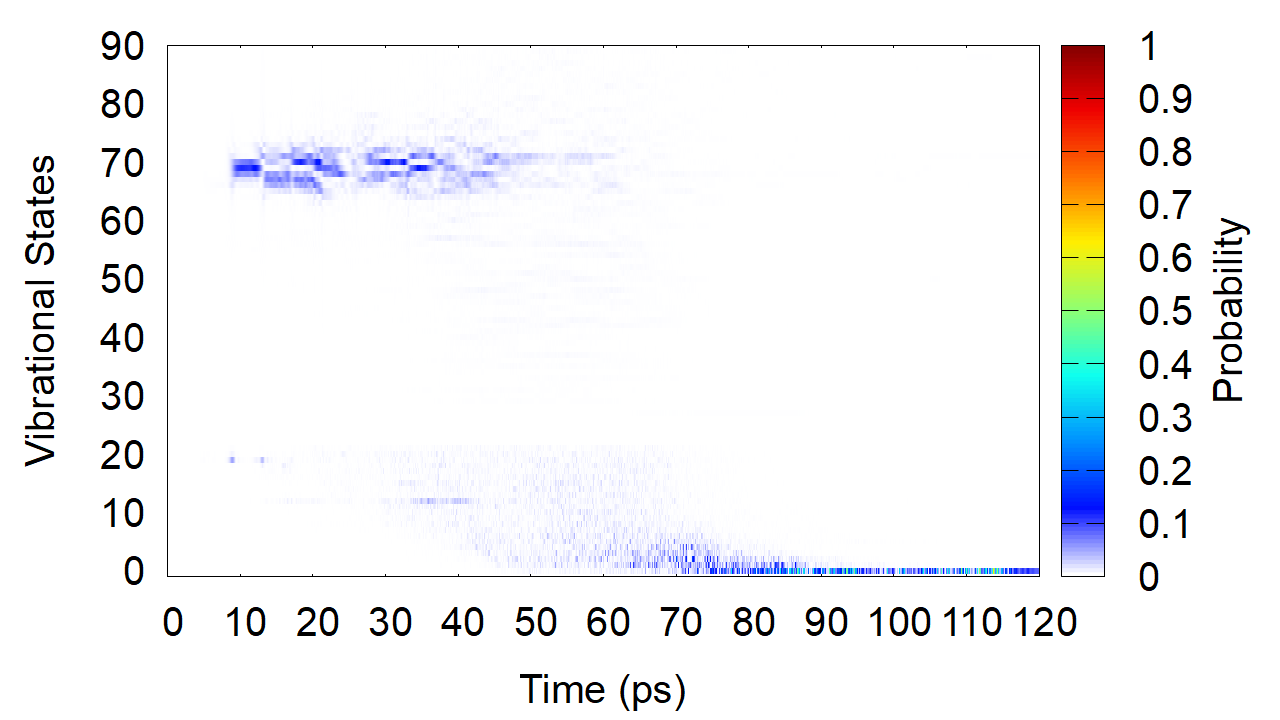}
     \caption{}
     \label{Fig10c}
 \end{subfigure}
 \hfill
 \begin{subfigure}[h]{8cm}
     \centering
     \includegraphics[width=\textwidth]{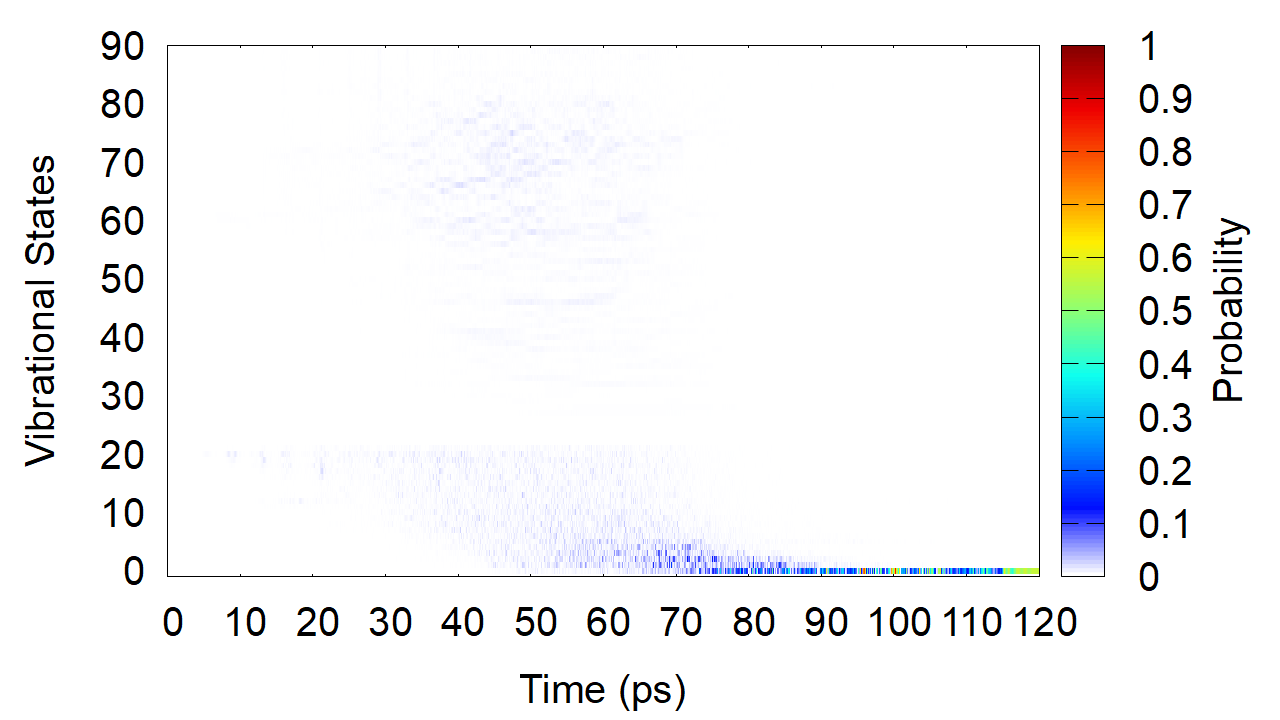}
     \caption{}
     \label{Fig10d}
 \end{subfigure}
  \hfill
 \begin{subfigure}[h]{8cm}
     \centering
     \includegraphics[width=\textwidth]{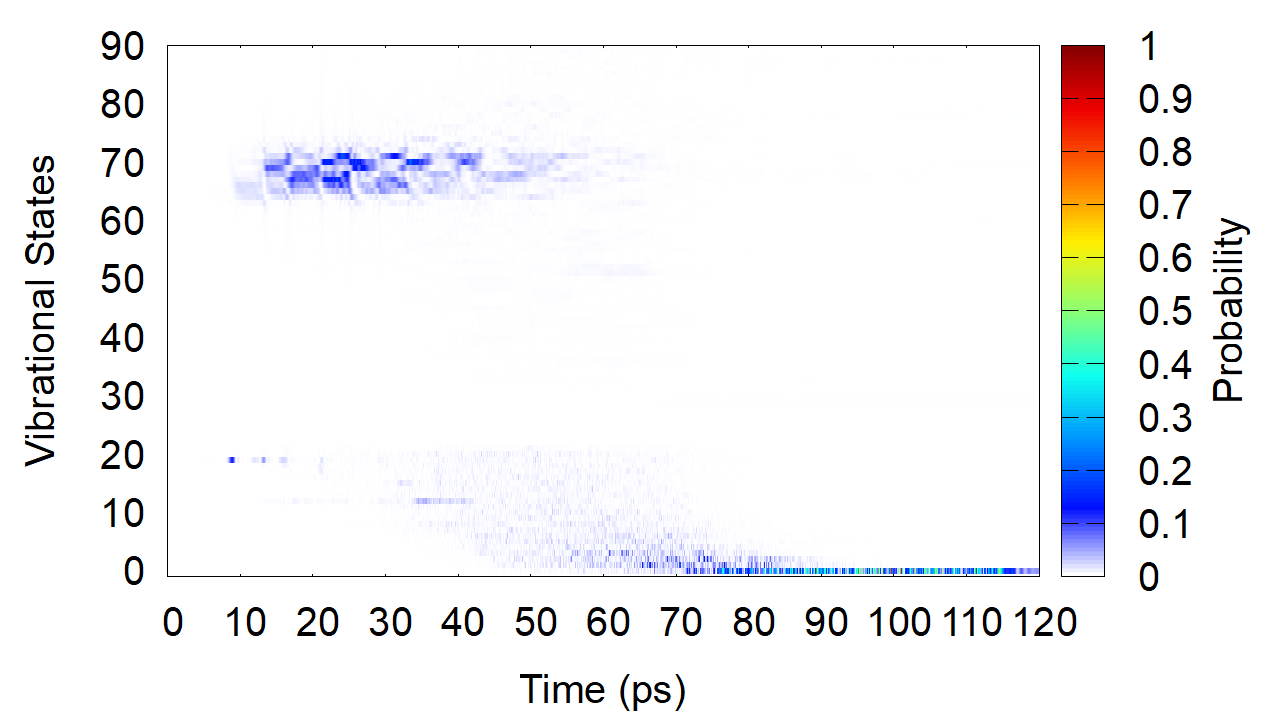}
     \caption{}
     \label{Fig10e}
 \end{subfigure}
    \caption{Population dynamics of the vibrational levels for the angular momentum: (a) $l=0$. (b) $l=1$. (c) $l=2$. (d) $l=3$. (e) $l=4$.}
    \label{Fig10}
\end{figure}

\clearpage
Panels (a)-(e) of Figure \ref{Fig10} show the population dynamics of the vibrational states for each value of angular momentum. In initial state $t=0$ the initial Gaussian wavepacket is located at $l=0$ and between the vibrational states $n\in(60,80)$, in the time interval $t\in(20,60)$, the system presents a complex behavior where there are free-free transitions as well as free-bound and bound-bound transitions between all rotational sublevels, for $t\in (60,100)$, essentially all population is in the bound energy levels, in the interval $t\in(100,120)$ the population is restricted to $\nu=0$ vibrational states. Considering Fig. \ref{Fig9} and \ref{Fig10}, we note that between $t=60$ and $t=118$, the population of the system undergoes transitions between vibrational levels of bound states, thus, according to the selection rules for the rotational transitions, for each transition, the population is forced to change its rotational state. This fact explains the complex transitions observed among the rotational levels and also why photoassociation alone is capable of producing some degree of orientation.

We now consider the photoassociation in a set of bound states. Thus, we perform the photoassociation of the initial wavepacket in the ten lowest vibrational bound states of the system, that is, from $\nu=0$ to $\nu=9$, so that the objective functional is given by, 

\begin{equation} \label{Eq.27}
J=  \langle \psi(t_f) |\boldsymbol{\hat O}|  \psi(t_f) \rangle  =\sum^{9}_{\nu=0}\sum^{4}_{l=0} \bra{\nu,l} \psi(t_f) \rangle |^2.
\end{equation}.

\begin{figure} [h]
    \centering
    \includegraphics[width=11cm]{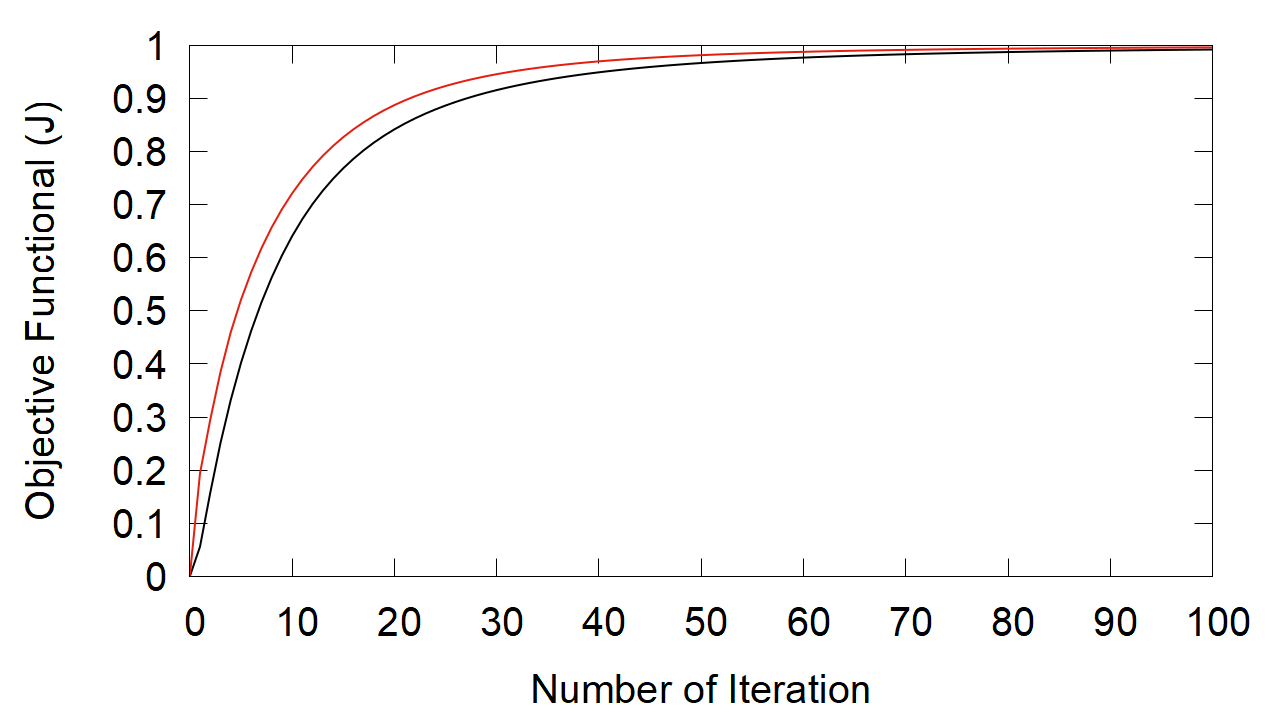}
    \caption{Optimization of the cost functional for each iteration of the TBQCP algorithm, the black line shows photoassociation for state $\nu=0$ and the red line for $\nu=[0,9]$.}\label{Fig11}
\end{figure}

Figure \ref{Fig11} shows a comparison between the TBQCP iterations for the objective functional when the photoassociation of the initial condition is performed at $\nu=0$ (black line) and at $\nu\in(0,9)$ (red line). Maximizing the functional requires a smaller number of iterations of the algorithm when considering the first ten lowest-energy vibrational levels compared to just the $\nu=0$ subspace. We can note that the cost functional reaches $0.8$ in the $10$th iteration for black curve and in the $20$th iteration for red curve, and their maximum values in the $80$th and $100$th iterations, respectively.

\begin{figure} [h]
    \centering
    \includegraphics[width=11cm]{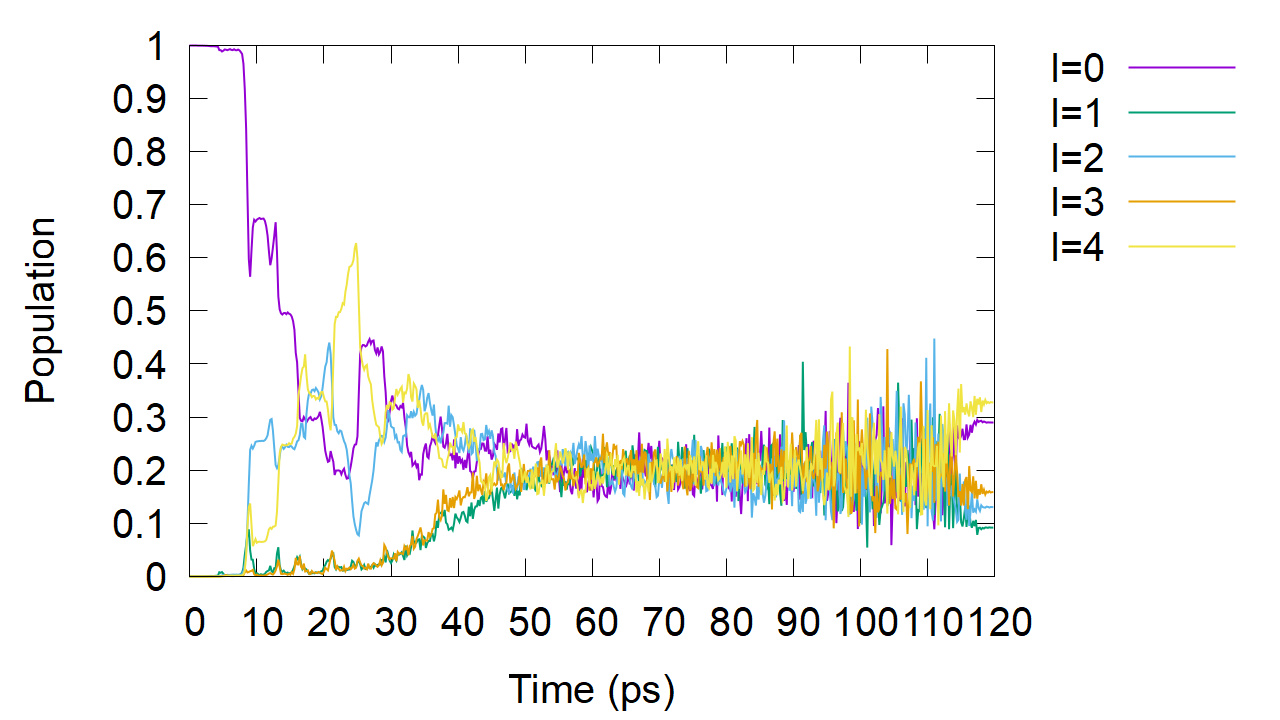}
    \caption{ Dynamic population for rotational levels $l=0,1,2,3,4$.}\label{Fig12}
\end{figure}

 Figure \ref{Fig12} shows the population dynamics for each value of the angular momentum. At the end of the control field the probability is distributed as follows: $\sim 0.28$ in $l=0$; $\sim 0.09$ in $l=1$; $\sim 0.14$ in $l=2$; $\sim 0.16$ in $l=3$; and $\sim 0.33$ in $l=4$, this distribution corresponds to a molecular orientation of $0.12$. Performing photoassociation in a set of bound vibrational states has a certain computational gain since a smaller number of iterations of the algorithm are required to reach the maximum value of the cost functional. However, it can lead to a smaller orientation of the molecule, in comparison to the photoassociation in a single vibrational state.


\subsection{Simultaneous Photo-association and Molecular Orientation}

In this section our objective is to drive the initial unbound wave function represented by Eq. \eqref{eq:initial_wp} to the $\nu=0$ rovibrational levels so that the population between the rotational levels has the optimal distribution that was found in section IIIa, as we wish to perform simultaneous control of both processes, the desired target operator  $\boldsymbol{\hat O} = \hat B$ is described by Equation \eqref{Eq.17}, the projection operator is $P_B=\sum P_B^l$ with $\nu=0$ and the molecular orientation operator is $\hat{A} = \cos (\theta)$, therefore, the cost functional is given by.

\begin{equation} \label{Eq.28}
J=  \langle \psi(t_f) |\boldsymbol{\hat O} | \psi(t_f) \rangle =\sum^{l_{max}}_{l=0}   \langle \Psi(t_f) | 0l \rangle \langle 0l|\cos(\theta)|0l \rangle   \langle 0l |\Psi(t_f) \rangle. 
\end{equation}

\begin{figure} [h]
    \centering
    \includegraphics[width=11cm]{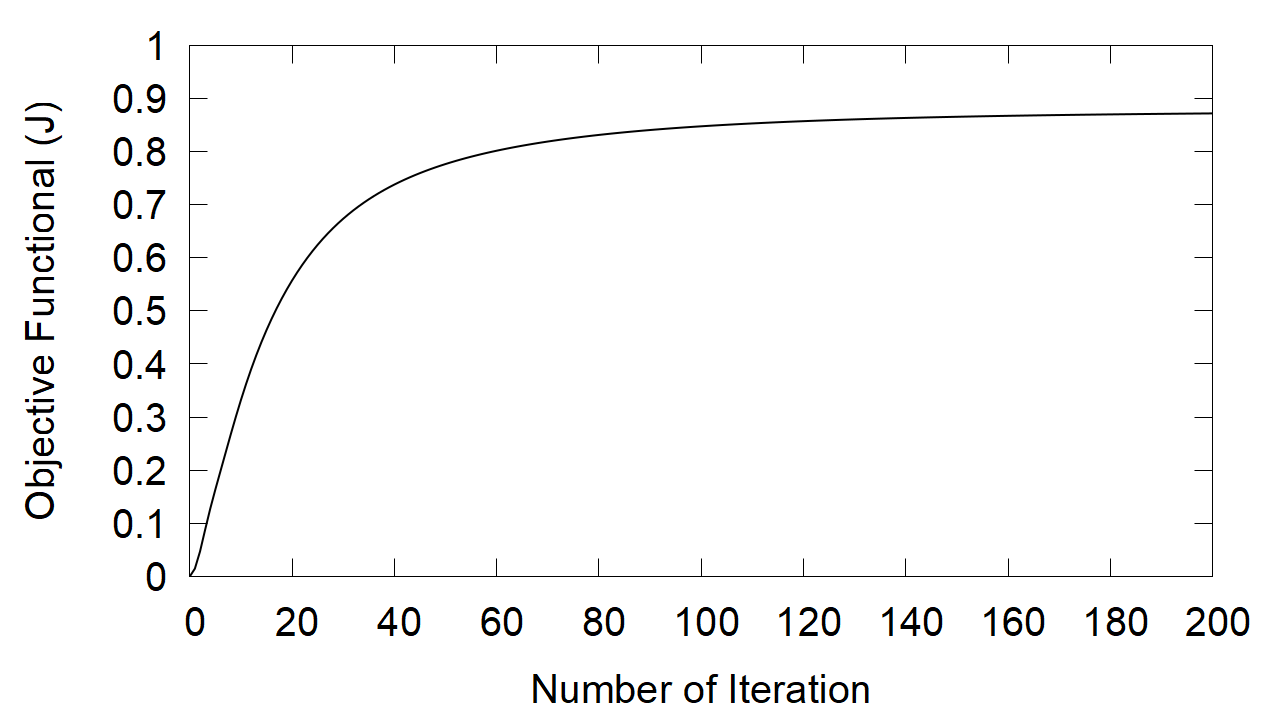}
    \caption{Optimization of the cost functional for each iteration of the TBQCP algorithm for the system $l_{max}=4$}
    \label{Fig13}
\end{figure}

Figure \ref{Fig13} shows the TBQCP iterations for the objective functional for the system with $l_{max}=4$, the initial value of $J$ is zero, between the 1st and 80th iteration the functional has significant growth reaching the value of $J\sim0.80$, then it has a small variation between the 81st and 200th iterations stabilizing at the value of $J\sim0.87$ at the end of the iterations.

\begin{figure} [h]
 \centering
 \begin{subfigure}[h]{11cm}
     \centering
     \includegraphics[width=\textwidth]{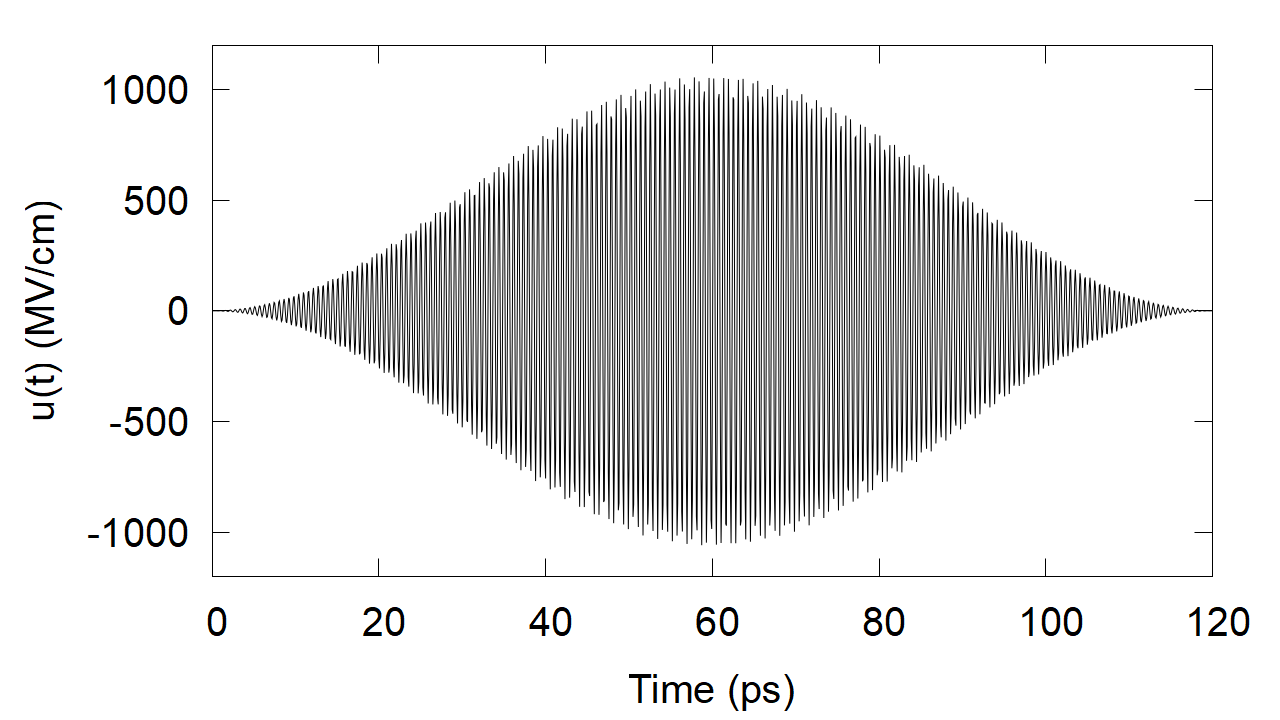}
     \caption{}
     \label{Fig14a}
 \end{subfigure}
 \hfill
 \begin{subfigure}[h]{11cm}
     \centering
     \includegraphics[width=\textwidth]{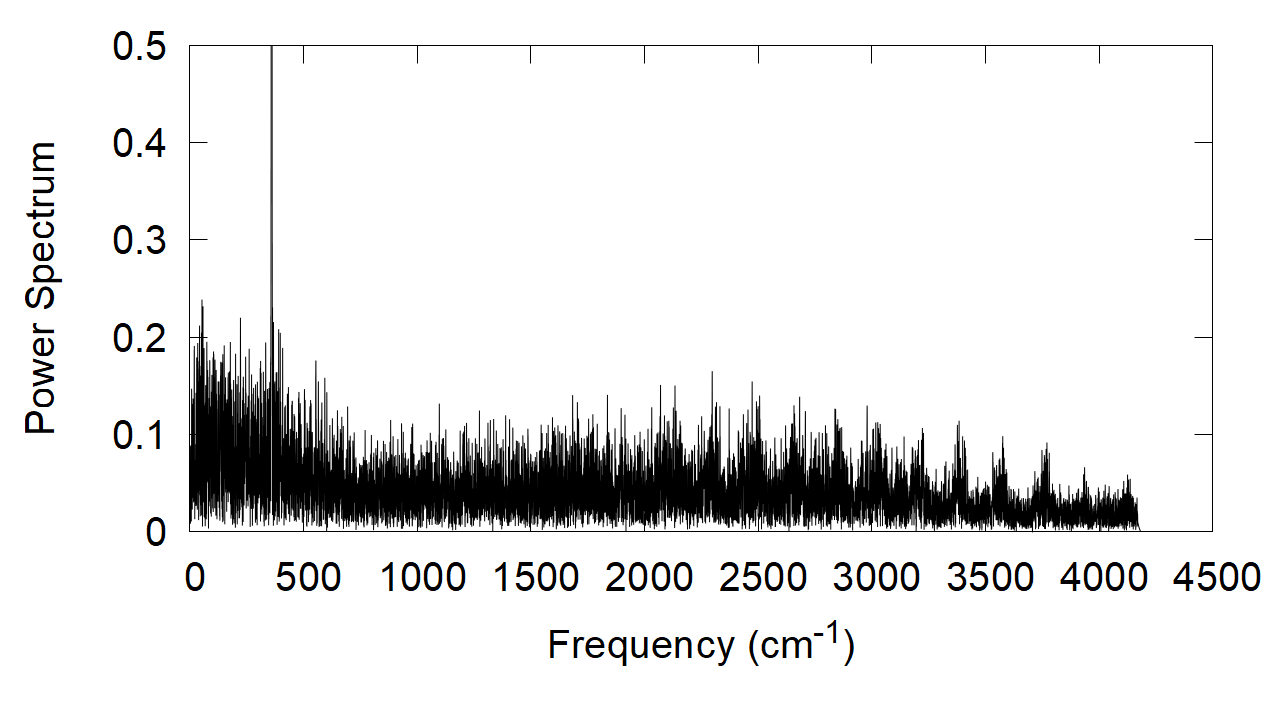}
     \caption{}
     \label{Fig14b}
 \end{subfigure}
 \hfill
    \caption{(a) Electric field resulting in the 200th iteration of the optimization algorithm. (b) The Fourier transform of the optimal field.}\label{Fig14}
\end{figure}

Note that in this case the cost functional does not reach the value of $J=1$ because, as shown previously, controlling the molecular orientation imposes a considerable number of rotation levels for its optimization to be more effective. Figure \ref{Fig14a} shows the electric field resulting from optimization and Figure \ref{Fig14b} its corresponding power spectrum. The optimized control field is composed by the trial field frequency, $\omega_c$, and a range of frequencies between $\omega=0$ and $\omega=4170$. Note that frequencies with values smaller than $\omega_c$ are more predominant than the others, with frequencies located in $(0,359)$ having power spectrum between 0 and 0.25, while those in the interval $(361,4170)$ vary between 0 and 0.18.

Comparing Figures \ref{Fig8} and \ref{Fig14}, we note that the power spectrum of the optimal field referring to the simultaneous control of photoassociation and molecular orientation varies between $0$ and $0.25$, while for the control of only photoassociation the variation is from $0$ to $0.16$. Furthermore, there is a frequency region  $\omega\in(2743,4170)$ that is present only in the optimal field represented by Figure \ref{Fig14a}. 

\begin{figure} [h]
 \centering
 \begin{subfigure}[h]{11cm}
     \centering
     \includegraphics[width=\textwidth]{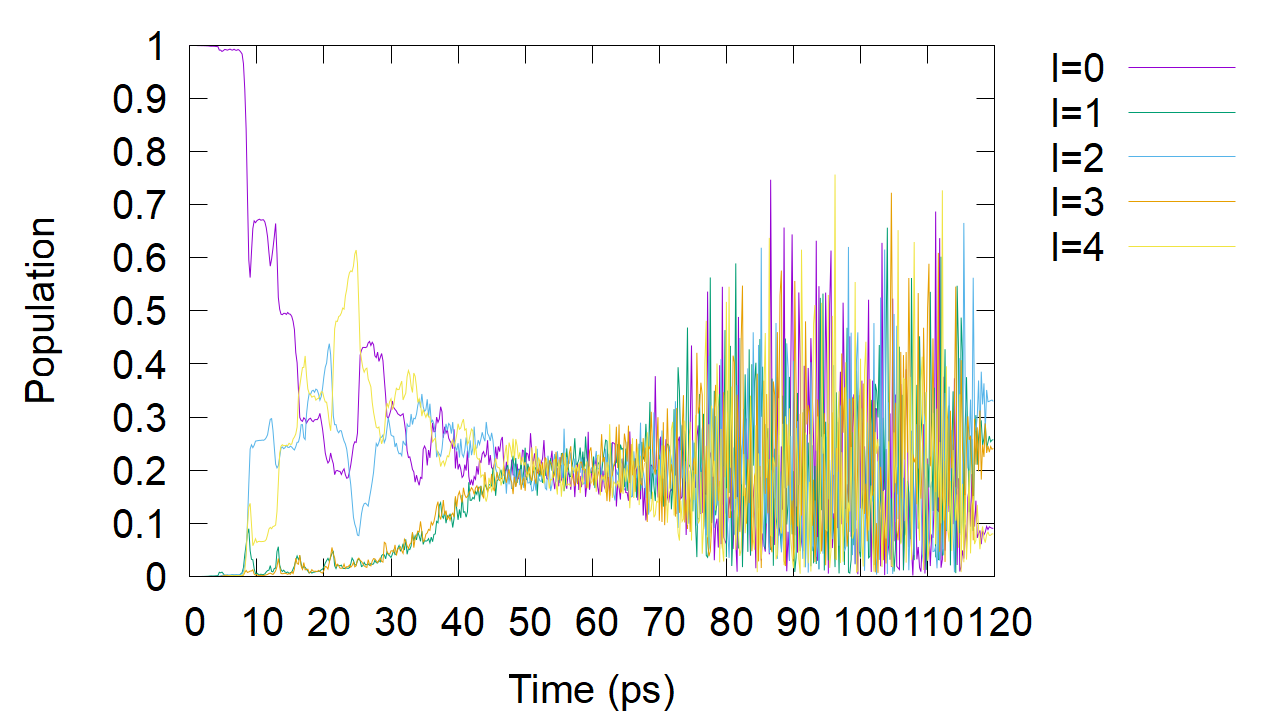}
     \caption{}
     \label{Fig15a}
 \end{subfigure}
 \hfill
 \begin{subfigure}[h]{11cm}
     \centering
     \includegraphics[width=\textwidth]{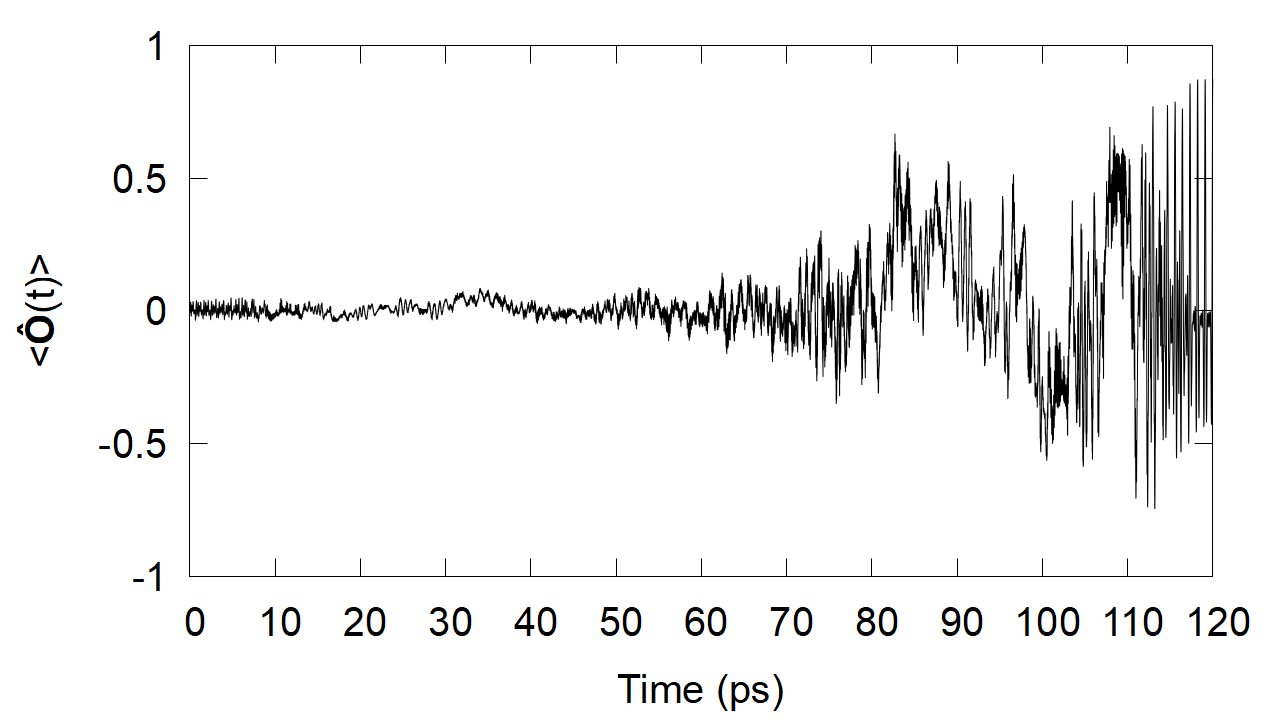}
     \caption{}
     \label{Fig15b}
 \end{subfigure}
 \hfill
    \caption{(a) Dynamic population for rotational levels $l=0,1,2,3,4$. (b) Expected value of operator $ \boldsymbol{\hat O}=\hat{B}$  in time} \label{Fig15}
\end{figure}

Figure \ref{Fig15a} shows the population dynamics between the rotation levels of the system. It is seen that the population distribution has a behavior similar to the dynamics of the photoassociation control of Figure \ref{Fig9a}, for $t=0$ the initial population is at $l=0$ and begins to distribute between the rotation levels around $t=8$ and presents a complex dynamics between $t=8$ and $t=114$. At $t=115$ the molecule begins to be oriented, reaching the optimal distribution of: $\sim 0.09$ in $l=0$ and $l=4$, $\sim 0.26$ in $l=1$, $\sim 0.32$ in $l=2$ and $\sim 0.24$ in $l=3$ at the end of the action of the optimized electric field.

\begin{figure} [h]
 \centering
 \begin{subfigure}[h]{8cm}
     \centering
     \includegraphics[width=\textwidth]{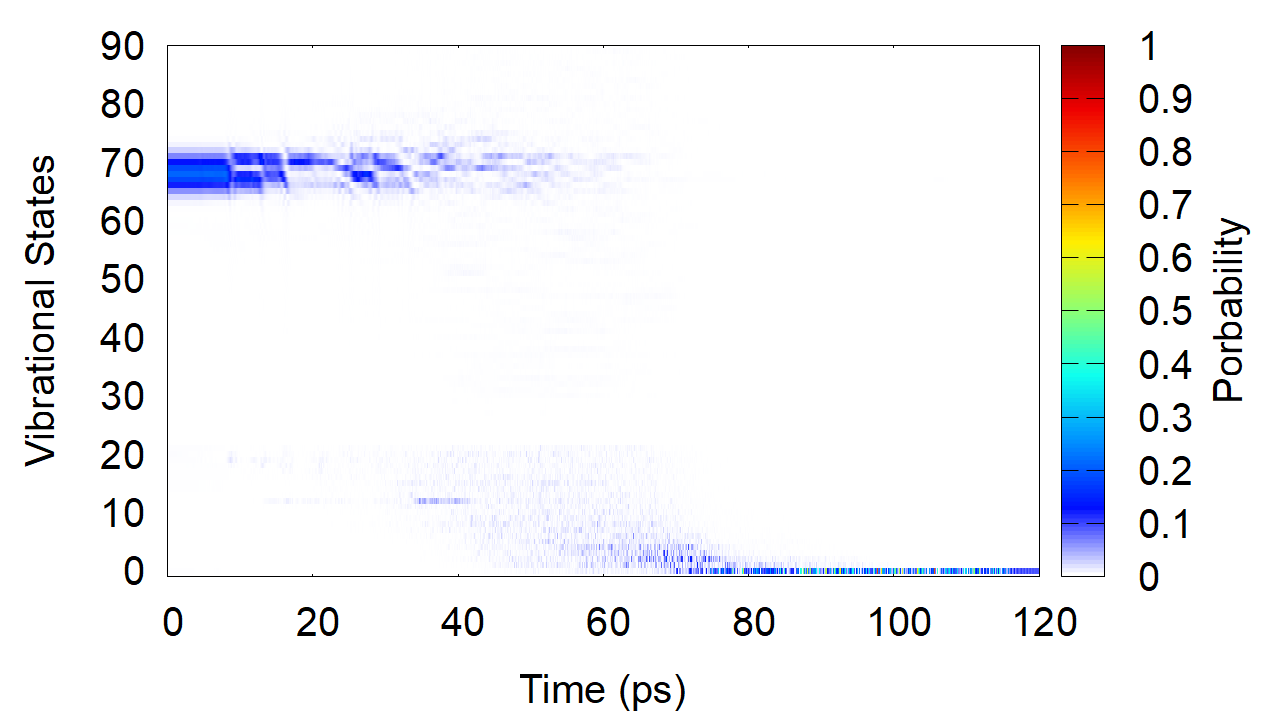}
     \caption{}
     \label{Fig16a}
 \end{subfigure}
 \hfill
 \begin{subfigure}[h]{8cm}
     \centering
     \includegraphics[width=\textwidth]{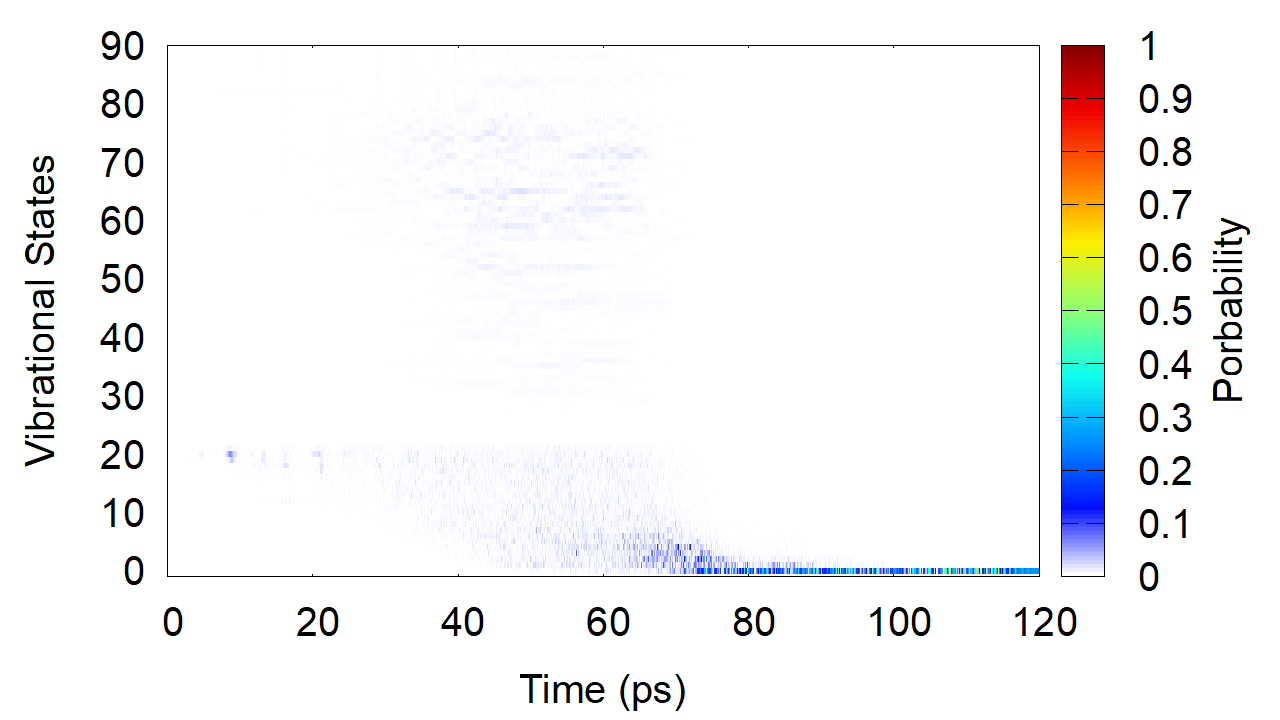}
     \caption{}
     \label{Fig16b}
 \end{subfigure}
 \hfill
 \begin{subfigure}[h]{8cm}
     \centering
     \includegraphics[width=\textwidth]{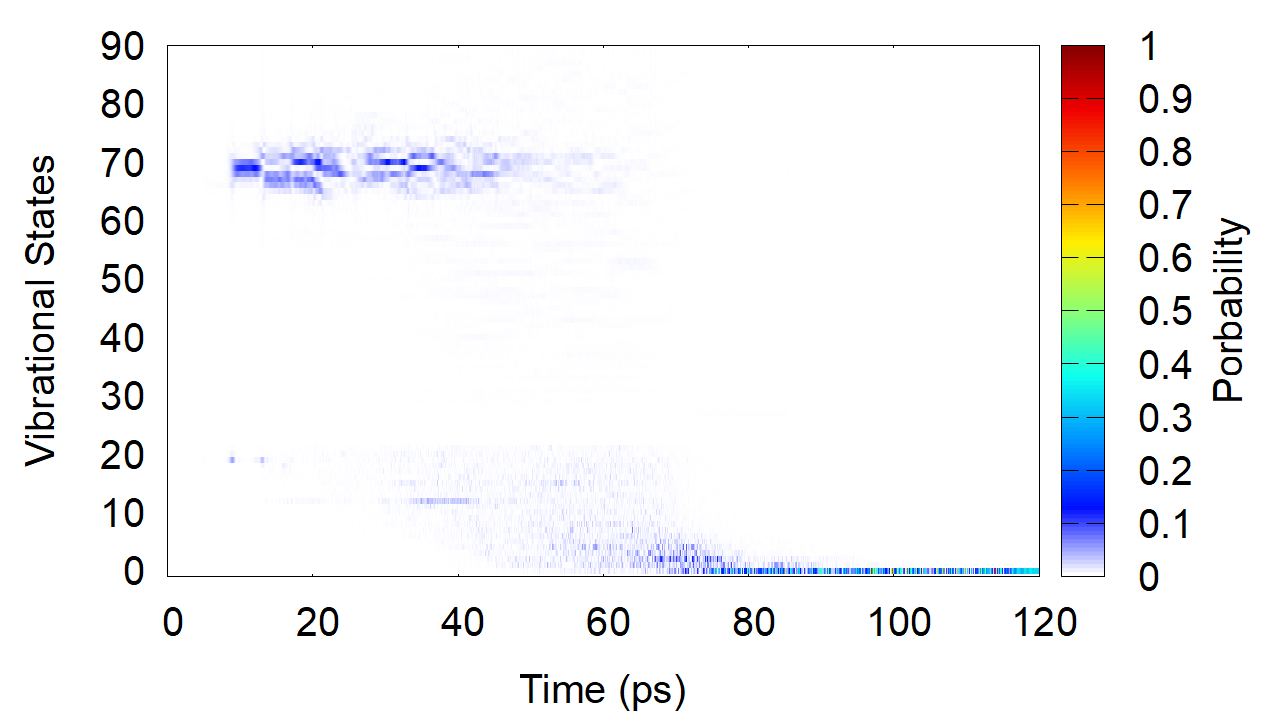}
     \caption{}
     \label{Fig16c}
 \end{subfigure}
 \hfill
 \begin{subfigure}[h]{8cm}
     \centering
     \includegraphics[width=\textwidth]{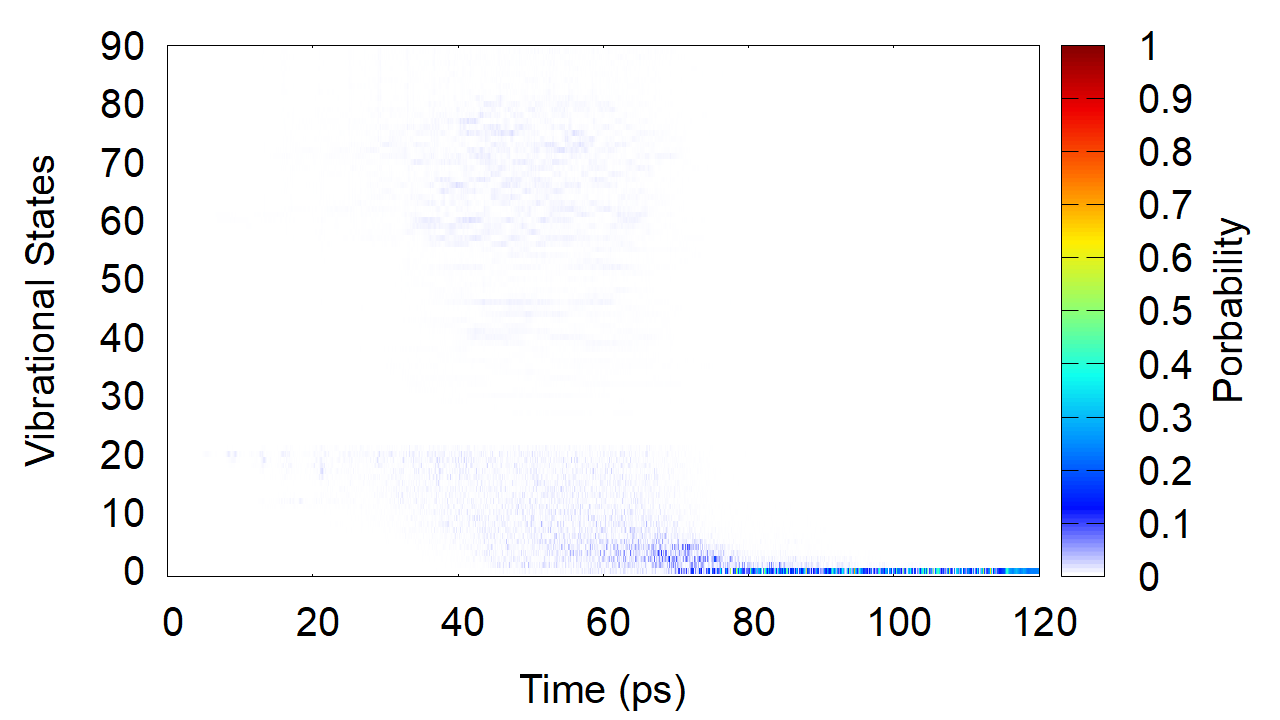}
     \caption{}
     \label{Fig16d}
 \end{subfigure}
  \hfill
 \begin{subfigure}[h]{8cm}
     \centering
     \includegraphics[width=\textwidth]{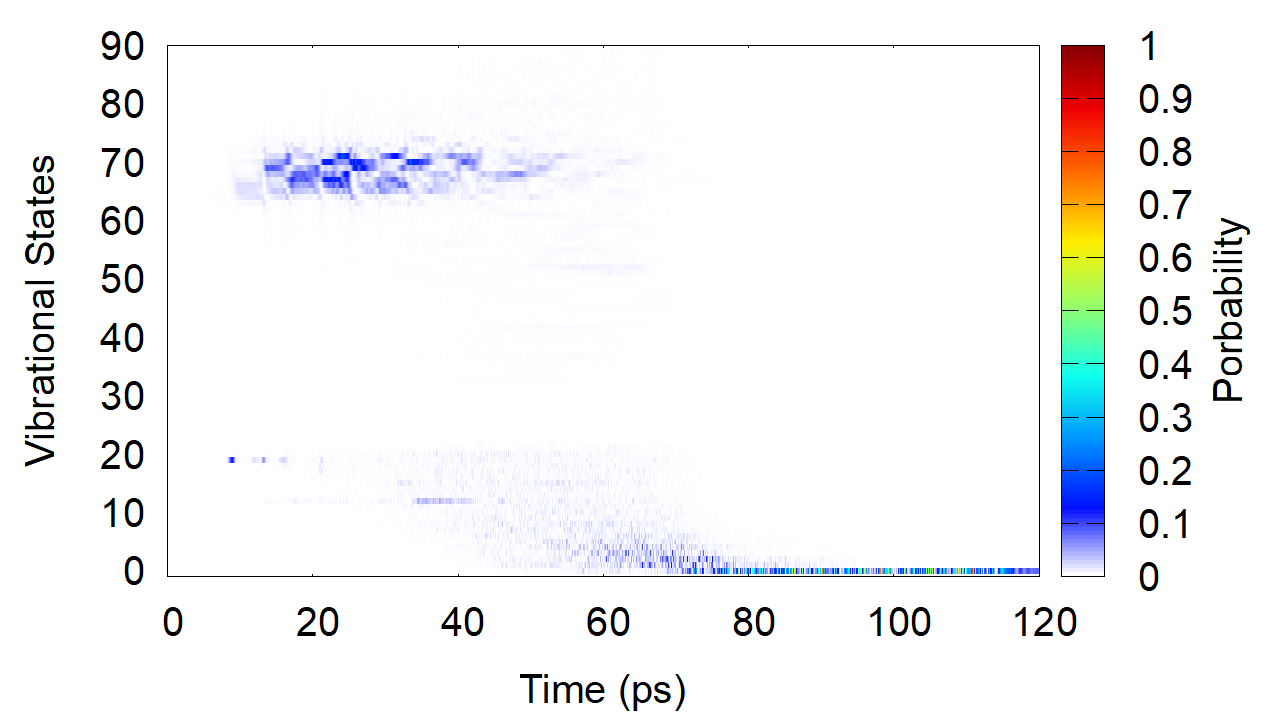}
     \caption{}
     \label{Fig16e}
 \end{subfigure}

    \caption{ Dynamic population of vibrational level for the rotation degree in (a) $l=0$. (b) $l=1$. (c) $l=2$. (d) $l=3$. (e) $l=4$.}
    \label{Fig16}
\end{figure}

Figure \ref{Fig15b} shows the expectation value of the operator $\boldsymbol{\hat O}$ over time for the optimized electric field. In the interval $t\in(0,60)$, the system presents a small variation around zero, between $t=60$ and $t=118$ the observable begins to have positive and negative values that vary between amplitudes of -0.8 and 0.8, at the end of the interaction with the optimal field the average value of the operator is 0.87.

Figure \ref{Fig16} shows the population dynamics for the initial state, in $t=0$ the Gaussian wavepacket is located at the vibrational levels $n\in(65,75)$. For $t\in(20,60)$, there is a complex dynamic involving many transitions among rovibrational levels, including transitions from unbound states to bound states (photoassociation), among unbound states and also among bound states (vibrational cooling). In the interval $t\in(60,100)$ essentially all the population is photoassociated in bound rovibrational levels. In the interval $t\in(110,119)$, the population is constrained to the $\nu=0$ levels of the system but without the desired orientation. In the final time, $t=120$ the population for $\nu =0$ and $l=0,1,2,3,4$ is distributed in the best possible way. Comparing Figures \ref{Fig15} and \ref{Fig16}, the photoassociation starts to increase in the interval $t\in(40,100)$, after this time the population is in the ground state and the control field begins to orient the molecule for $t\in(100,120)$. Showing that it is possible to perform the photoassociation directly in the ground vibrational state and orient the molecule formed through the simultaneous control of these two processes.

\section{Conclusion}

In this study, we addressed the problem of forming an oriented polar molecule from a pair of colliding atoms by means of linearly polarized laser pulse. We illustrated the situation with a model system for the O + H collision pair interacting with an infrared pulse. Our goal was to explore whether a single shaped pulse could simultaneously accomplish three process: photoassociation, vibrational stabilization, and molecular orientation. All of the dynamics is carried out in the electronic ground state thanks to the permanent dipole associated to the involved transitions. By employing the TBQCP optimization algorithm, we demonstrated that optimized fields can indeed perform these tasks in a unified manner. To this end, we set as our target observable the restriction of the orientation operator to a set o bound states.

We showed that photoassociation and vibrational stabilization together can lead to a partial orientation of the molecule, which is consistent with earlier reports. Since the coupling of the unbound states is non-negligible only with excited vibrational states, in the combination of photoassociation and vibrational stabilization processes, the control field first transfers population into higher-lying vibrational levels and subsequently drives it toward the lowest vibrational states. During this vibrational cooling, the system naturally has to occupy several rotational levels due to the transition section rule $l\rightarrow l\pm 1$. When orientation is explicitly incorporated into the control objective, the control field substantially enhances the orientation, organizing the rotational population to yield optimal orientation.  This result confirms that a single control pulse is in principle capable of guiding the entire process from initially unbound atoms to a stable oriented polar molecule.

\section{Acknowledgments}

This research was supported by resources supplied by FAPESP-São Paulo Research Foundation (No. 2024/09015-5) and by the Center for Scientific Computing (NCC/GridUNESP) of the São Paulo State University (UNESP).
E.D.L. acknowledges support from Brazilian agencies CNPq (No. 301318/2019-0, 304398/2023-3) and FAPESP (No. 2019/14038-6 and No. 2021/09519-5). E.F.L. acknowledges support from FAPESP (No. 2024/22593-8).


%

\end{document}